\documentclass[a4paper,11pt]{article}
\pdfoutput=1 

\usepackage{jcappub} 

\usepackage[T1]{fontenc} 
\usepackage{textcomp}

\usepackage{pgfplots}
\pgfplotsset{compat=1.15}
\usepackage{mathrsfs}
\usetikzlibrary{arrows}

\title{\boldmath T-RAX: Transversely Resonant Axion eXperiment}

\author{Chang Lee,\note{Corresponding author.}}
\author{Olaf Reimann}
\affiliation{Max-Planck-Institut f\"ur Physik (Werner-Heisenberg-Institut),\\ F\"ohringer Ring 6, 80805 M\"unchen, Germany}

\emailAdd{changlee@mpp.mpg.de}

\abstract{We propose to use an elongated rectangular waveguide near its cutoff frequency for axionic dark matter searches. The detector's large surface area allows for significant signal power, while its narrow transverse dimension and tapered-waveguide coupling suppress parasitic modes. The proposed system can fit inside a solenoid magnet and is sensitive to the QCD-axion in the axion mass $40-400\,\mu$eV. We describe the theoretical principles of the new design, present simulation results, and discuss the implementation.}

\begin{document}
\maketitle
\flushbottom

\section{Introduction}
While numerous astronomical observations strongly support the existence of dark matter~\cite{Planck2018}, its identity remains one of the biggest mysteries of modern physics. Axions were initially motivated to solve the Strong CP problem~\cite{PhysRevD.16.1791, PhysRevLett.38.1440,PhysRevLett.40.223,PhysRevLett.40.279}, but also turned out to be excellent candidates for dark matter~\cite{PRESKILL1983127,ABBOTT1983133,DINE1983137}. Axions in the mass range 40--400$\,\mu$eV are particularly favored in the scenarios where the Peccei-Quinn symmetry breaks after cosmic inflation~\cite{Borsayini,Dine:2017swf,Klaer:2017ond,Buschmann:2019icd,Buschmann:2021sdq,Ballesteros:2016euj}. 
    
Unfortunately, most existing axion dark matter searches are not optimized in this mass range. They utilize axion-photon coupling resonantly enhanced inside a cavity~\cite{PhysRevD.32.2988, PhysRevD.36.974, PhysRevLett.120.151301, PhysRevLett.118.061302}. While the strategy has been extraordinarily successful in the mass below $\mathcal{O} (10)~\mu$eV, it becomes rapidly inefficient as the mass increases. The cavity's volume shrinks inversely proportional to the axion mass. The shrinking volume reduces the signal power and increases the search time. 

Dielectric haloscopes are promising alternatives. They detect axion-induced traveling waves from dielectric interfaces~\cite{PhysRevLett.118.091801, orpheus}. The traveling waves resonate only in the propagating direction and not in the transverse dimensions. Their signal power can increase proportionally to the area of the dielectrics, independent of the axion mass. Wave detection using a horn antenna or tapered waveguide can couple to the exact transverse mode of interest, not the unwanted higher-order modes. Unfortunately,  the larger dielectric also requires a larger magnet. In addition, the traveling wave detection system also requires a long optical path perpendicular to the magnetic field. As a result, the dielectric haloscopes need a large custom dipole magnet.

Another recent proposal suggests using a \emph{thin-shell cavity}: two parallel conducting plates separated by one half wavelength~\cite{Kuo_2020,Kuo_2021}. The plates support only the fundamental transverse modes. The cavity's volume can increase independent of the frequency in two other dimensions. The design varies the separation width to scan a wide  axion mass range. The detector coherently sums the signals from multiple ports to filter out the unwanted high-order modes. The system can also fit in a simple solenoid magnet. We point out two challenges of the design. First, the tuning mechanism allows multiple degrees of freedom and is mechanically complicated. Slight misalignment or tilt of the mechanism splits or shifts the peaks. Synchronized operation of multiple such cavities will be even more challenging. Second, the plates still host multiple vertical higher-order modes, TE/TM $1ml$ and $0ml$, and identifying them with dipole antennas becomes rapidly challenging. These unwanted modes overlap with the desired modes and limit vertical scale-up. In addition, coherent signal summing is not proven in a resonant setup.

In this paper, we propose a novel axion detector concept that takes advantage of both the \emph{thin-shell cavity} and the dielectric haloscope. Our proposed Transversely Resonant Axion eXperiment (T-RAX) utilizes an elongated rectangular waveguide as a large-volume resonator with few higher-order modes. Our approach adds dielectric slabs to induce longitudinal resonance and to tune the system. To collect the axion-induced signal, T-RAX uses a tapered waveguide, which is relatively simple and easy to implement. T-RAX can potentially sense some QCD-axion models in the centimeter-wave range, but it is also relatively small and can fit inside a solenoid magnet. In the next section, we discuss the principle of T-RAX. In Sec.\,\ref{sec:simulation}, we present the result from finite element method simulations. In Sec.\,\ref{sec:packaging}, we discuss how to reduce the detector profile to fit it inside a solenoid magnet. In Sec.\,\ref{sec:sensitivity}, we discuss the projected sensitivity of T-RAX. Finally, we summarize and conclude in Sec.\,\ref{sec:conclusion}.

\section{Theory}\label{sec:theory}
In this section, we develop a simple model of T-RAX, a dielectric haloscope operating near a waveguide's cutoff frequency. First, we show how the axion induces a strong transverse resonance between two parallel plates. Then we discuss how this resonating standing wave generates a traveling wave from a shorted end. Finally, we will see how an additional longitudinal reflection can further boost the traveling wave power.

\subsection{Transverse resonance of parallel plates}

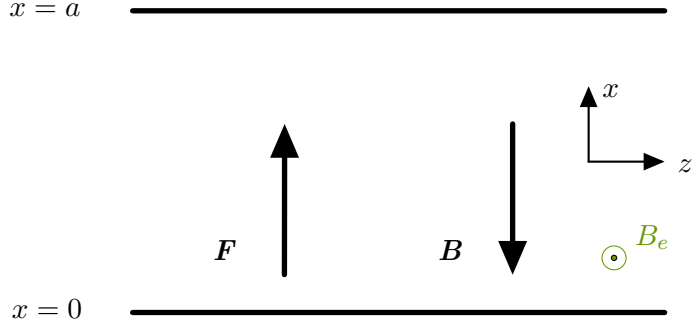
\begin{figure}
\begin{center}
\definecolor{wwzzqq}{rgb}{0.4,0.6,0.}
\begin{tikzpicture}[line cap=round,line join=round,>=triangle 45,x=1.0cm,y=1.0cm]
\clip(-2.881187063715945,-1.297361575133861) rectangle (9.405285121446205,5.680957754796707);
\draw [line width=2.pt] (0.,0.)-- (7.,0.);
\draw [->,line width=2.pt] (1.997178880280643,0.5019830980035995) -- (2.,2.5);
\draw [->,line width=2.pt] (5.,2.5) -- (5.,0.5);
\draw (0.9221623341135229,1.1004021756716662) node[anchor=north west] {\textit{\textbf{F}}};
\draw (3.8821603437286307,1.1004021756716662) node[anchor=north west] {\textit{\textbf{B}}};
\draw [line width=2.pt] (0.,4.)-- (7.,4.);
\draw (-1.7567185460967982,4.225763202583698) node[anchor=north west] {$x=a$};
\draw (-1.7236459426374116,0.30665969264638815) node[anchor=north west] {$x=0$};
\draw [color=wwzzqq](6.478359715290486,1.2988377964279856) node[anchor=north west] {$B_e$};
\draw [->,line width=0.8pt] (6.,2.) -- (7.,2.);
\draw [->,line width=0.8pt] (6.,2.) -- (6.,3.);
\draw (6.0484158703184585,3.1509035901536344) node[anchor=north west] {$x$};
\draw (7.04059397410006,2.158725486372037) node[anchor=north west] {$z$};
\begin{scriptsize}
\draw [color=wwzzqq] (6.334204199877346,0.7239749730973842) circle (4.5pt);
\draw [fill=wwzzqq] (6.334204199877346,0.7239749730973842) circle (1.0pt);
\end{scriptsize}
\end{tikzpicture}
\caption{\label{fig:TransverseResonance} Diagram for the transverse resonance.}
\end{center}
\end{figure}

Fig.~\ref{fig:TransverseResonance} displays two infinite conducting plates at $x=0$ and $a$ in free space with a wave impedance $Z_0$. A strong external DC magnetic field $B_e$ runs parallel to the plates ($y$-axis). The axion field $a$ couples with $B_e$ and induces an electric field in the vacuum parallel to the plates. The electric field oscillates with frequency $\nu_a = \frac{m_a}{h}$, where $m_a$ is the axion mass and $h$ is the Planck constant. The electric field has an amplitude 
\begin{equation}
    E_0 = \frac{\alpha}{2\pi\varepsilon_0} C_{a\gamma} B_e \theta \simeq 1.3 \times 10^{-12} \text{\,[V/m]} \frac{B_e}{10\text{\,T}} |C_{a\gamma}|.
\end{equation}
Here, $\alpha$ is the fine structure constant, $\varepsilon_0$ is the vacuum electric permittivity, $C_{a\gamma}$ is the model-dependent dimensionless axion-photon coupling, and $\theta$ is the axion field. We also assume that the axion is responsible for all dark matter density.

$E_0$ excites two transversely travelling waves moving forward and backward, along the $x$-axis. They have complex amplitudes $F$ and $B$ at $x=0$ and a transverse wave number $k_x$. We calculate $F$ and $B$ by applying the following boundary conditions at the plates:
\begin{equation}
\begin{split}
    F + B + E_0 &= 0 \text{ at } x=0\text{, and} \\
    F e^{-ik_xa} + B e^{ik_xa} + E_0 &= 0 \text{ at } x=a.
\end{split}
\end{equation}
The equations above yields
\begin{equation}
\begin{split}
    F &= - \frac{1}{1 + e^{-ik_xa}} E_0\text{ and}\\
    B &= - \frac{e^{-ik_xa}}{1+e^{-ik_xa}} E_0.
\end{split}
\end{equation}

The sum of $F$ and $B$ forms a standing wave. The electric field  $E^{tr}_y$ at the center of the waveguide ($x=\frac{a}{2}$) is
\begin{equation}
\begin{split}
    E^{tr}_y &= F e^{-ik_x a/2} + B e^{ik_x a/2} \\
    &= - \frac{E_0}{1 + e^{-ik_x a}} (e^{-ik_x a/2} + e^{-ik_x a/2})\\
    &= - \frac{2 E_0 }{e^{ik_x a/2} + e^{-ik_x a/2}} \\
    &= - \frac{E_0}{\cos{\frac{k_x a}{2}}}.
\end{split}\label{eq:e_tr_transverse}
\end{equation}
$|E^{tr}_y|$ diverges to infinity as $k_x a \to \pi$\footnote{$E_0$ also excites $k_x a = 3\pi, 5\pi, ...$ if the spacing between the plates is wider, \emph{i.e.} $E_0$ excites TE30 and TE50 modes. Unfortunately, extracting power from these modes is practically very challenging.}, at the TE10 mode cutoff frequency. The divergence is expected because the parallel plates form a loss-less cavity with an infinite quality factor. Unlike a cavity, however, $-\frac{1}{\cos{\frac{k_x a}{2}}} \gg 1$ even when $k_x a $ is slightly larger than $\pi$, or slightly above the cutoff frequency, and $|E^{tr}_y| \gg |E_0|$. Such a \emph{side-band resonance} is a unique characteristic of the axion-induced emission. $|E^{tr}_y|$ and the detectable signal power increases as $k_x a \to \pi$. Unfortunately, operation near the cutoff frequency also requires more precise machining and increases conduction loss. The optimal spacing between the plates should consider these factors.  

\subsection{Short-circuited waveguide}\label{sec:short-curcuited}
\begin{figure}
\begin{center}
\definecolor{wwzzqq}{rgb}{0.4,0.6,0.}
\begin{tikzpicture}[line cap=round,line join=round,>=triangle 45,x=1.0cm,y=1.0cm]
\clip(-1.1280694321920255,-1.1463779999968846) rectangle (9.331471323555956,3.531361060212609);
\draw [line width=2.pt] (0.,0.)-- (6.,0.);
\draw [line width=2.pt] (0.,3.)-- (6.,3.);
\draw [line width=2.pt] (0.,3.)-- (0.,0.);
\draw [line width=2.pt,dash pattern=on 3pt off 3pt] (6.,3.)-- (6.,0.);
\draw [shift={(5.300790267966324,1.5012341203518473)},line width=2.pt]  plot[domain=-1.5707963267948966:1.5707963267948966,variable=\t]({1.*0.5*cos(\t r)+0.*0.5*sin(\t r)},{0.*0.5*cos(\t r)+1.*0.5*sin(\t r)});
\draw [line width=2.pt] (5.300790267966324,2.0012341203518473)-- (4.800791012967033,2.00037098643753);
\draw [line width=2.pt] (5.300790267966324,1.0012341203518473)-- (4.800793602750105,0.9994079843600997);
\draw [->,line width=2.pt] (1.,2.) -- (2.4993856043411404,1.9986839172691884);
\draw [->,line width=2.pt] (2.5,1.) -- (1.,1.);
\draw [->,line width=2.pt] (7.,2.) -- (8.5,2.);
\draw (0.2520088619691664,2.282027025498272) node[anchor=north west] {$E_R$};
\draw (0.2665360019077052,1.366817209370328) node[anchor=north west] {$E_L$};
\draw (6.193609096831561,2.2384456056826556) node[anchor=north west] {$E_T$};
\draw (6.280771936462794,1.5266157486942546) node[anchor=north west] {$Z_L$};
\draw (4.392243744452742,1.773577127649414) node[anchor=north west] {$\Gamma$};
\draw (1.4432343369293532,1.7445228477723367) node[anchor=north west] {$\beta$};
\draw (-0.6922552340358596,-0.14400534423770742) node[anchor=north west] {$z=0$};
\draw (5.5689420794743905,-0.14400534423770742) node[anchor=north west] {$z=d$};
\draw (-0.663200954158782,2.354662725190966) node[anchor=north west] {$\rotatebox{90.0}{ \text{ mirror }  }$};
\draw [color=wwzzqq](3.171963989615478,0.625933072504849) node[anchor=north west] {$B_e$};
\draw [->,line width=0.8pt] (8.,0.) -- (9.,0.);
\draw [->,line width=0.8pt] (8.,0.) -- (8.,1.);
\draw (3.041219730168628,2.136755626112884) node[anchor=north west] {$Z$};
\draw (8.038555869025997,1.1198558304151682) node[anchor=north west] {$x$};
\draw (8.910184265338328,0.13201031459452978) node[anchor=north west] {$z$};
\draw (-0.7067823739743986,3.6330510397823805) node[anchor=north west] {$x=a$};
\begin{scriptsize}
\draw [fill=black,shift={(4.800793602750105,0.9994079843600997)},rotate=90] (0,0) ++(0 pt,5.25pt) -- ++(4.546633369868303pt,-7.875pt)--++(-9.093266739736606pt,0 pt) -- ++(4.546633369868303pt,7.875pt);
\draw [fill=wwzzqq] (2.9831111704144724,0.3499174136726118) circle (1.0pt);
\draw [color=wwzzqq] (2.9831111704144724,0.3499174136726118) circle (4.5pt);
\end{scriptsize}
\end{tikzpicture}
\caption{\label{fig:diagram} A simple T-RAX.}
\end{center}
\end{figure}
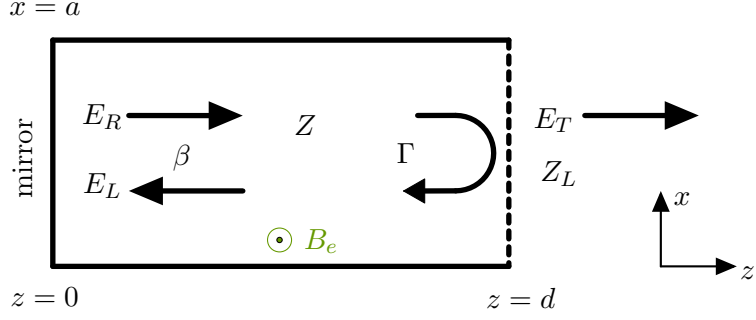

$E^{tr}_y$ induces a traveling wave from a conducting surface. We consider the geometry in Fig.~\ref{fig:diagram}, yet without the reflection $\Gamma$. A metallic mirror shorts the parallel plates at the left end ($z=0$). The parallel plates form a transmission line with a TE10-mode loss-less propagation constant $\beta$. The right and left traveling waves have complex amplitude $E_R$ and $E_L$ at the center of the mirror ($x=\frac{a}{2}$, $z=0$). The boundary condition here is\footnote{We ignore $E_0$ on the mirror because we are interested in the case where $|E^{tr}_y| \gg |E_0$|. Similarly, we ignore the axion-induced traveling wave at the interface to the load}
\begin{equation}
    E_R + E_L + E^{tr}_y = 0.
\label{eq:mirror_bc}
\end{equation}
Because $E_L=0$, $E_R = - E^{tr}_y = \frac{E_0}{\cos{\frac{k_x a}{2}}}$. The transverse resonance formed inside the parallel plates can source an axion-induced traveling wave stronger than $E_0$. 

\subsection{1D transmission line dielectric haloscope}\label{sec:1D_DH}
We introduce a load with an impedance $Z_L$ at $z=d$, where $d$ is larger than the free-space wavelength, and the transverse resonance is hardly disturbed at the load. Fig.~\ref{fig:diagram} shows the geometry. The reflection at the load is 
\begin{equation}
    \Gamma = \frac{Z_L-Z}{Z_L+Z},
\end{equation}
where $Z = \frac{k Z_0}{\beta}$ is the TE10 mode wave impedance inside the transmission line. Eq.~\ref{eq:mirror_bc} is still valid, but $E_L$ is no longer zero. 


The new boundary condition at $z=d$ is
\begin{equation}
    E_L e^{i\beta d} = \Gamma E_R e^{-i \beta d} + \Gamma E^{tr}_y.
\end{equation}
Solving the boundary conditions yields
\begin{equation}
    E_R =- \frac{1 + \Gamma e^{-i\beta d}}{1 + \Gamma e^{-2i\beta d}} E^{tr}_y.
\label{eq:longitudinal_boost}
\end{equation}
$E_R$ diverges to infinity when $\Gamma e^{-2i \beta d} \to -1$, \emph{i.e.} $E_R$ is maximum where $2\beta d + \angle \Gamma = \pi$. For example, if $\Gamma$ is a real minus number ($\angle \Gamma = \pi$), the system resonates at $\beta d = \pi$. We use this relation for tuning later. The resonance is similar to a TE101 mode, but $k_x$ is slightly larger than the cutoff wavenumber. A larger $|\Gamma|$ will yield a stronger longitudinal resonance and increase $E_R$. The operation near the cutoff frequency increases $|\Gamma|$: an empty waveguide's TE mode wave impedance diverges near the cutoff frequency, while that of a material-loaded waveguide is still close to its free space value.



A fraction of $E_R$ passes the reflective boundary and delivers a measurable power to the load. The transmitted wave is
\begin{equation}
    E_T = (1 + \Gamma) E_R e^{-i \beta d} = \frac{2Z_L}{Z_L + Z}E_R e^{-i \beta d}.
\end{equation}
The time-averaged power on the load is
\begin{equation}
\label{eq:leakagePower}
\begin{split}
    P_\text{sig} &= \frac{1}{2} \text{Re} \int_{A} E_T \times \frac{E_T^*}{Z_L} \\
    &= \frac{2 R_L}{|Z_L+Z|^2} \int_A |E_R|^2,
\end{split}
\end{equation}
where $R_L$ is the resistive component of $Z_L$: $R_L = \text{Re}(Z_L)$. $\int_{A}$ is the integration of the TE10 mode over the $xy$ cross-section of the waveguide. For a waveguide of width $a$ and height $b$, the integration yields $\frac{ab}{2}$. $\int_{A}$ and $P_\text{sig}$ scale inversely proportional to frequency $\nu$ ($\propto \nu^{-1}$) because the width of the waveguide has to shrink so that we operate near the cutoff frequency. The same is true for the example in Sec.~\ref{sec:short-curcuited}. For comparison, pure dielectric haloscopes use no transverse resonance, and its signal power is independent of the frequency ($\propto \nu^0$). 


We compare $P_\text{sig}$ to the axion-induced traveling wave power of a simple infinite mirror of the same cross-sectional area $ab$: $P_0 = \frac{E_0^2}{2 Z_0} ab$. We call this ratio a power boost.
\begin{equation}
    \frac{P_\text{sig}}{P_0} = \frac{2 Z_0 R_L}{|Z_L+Z|^2}\frac{|E_R|^2}{{E_0}^2}.
    \label{eq:power_boost}
\end{equation}

\begin{figure}
\centering 
\includegraphics[width=.6\textwidth]{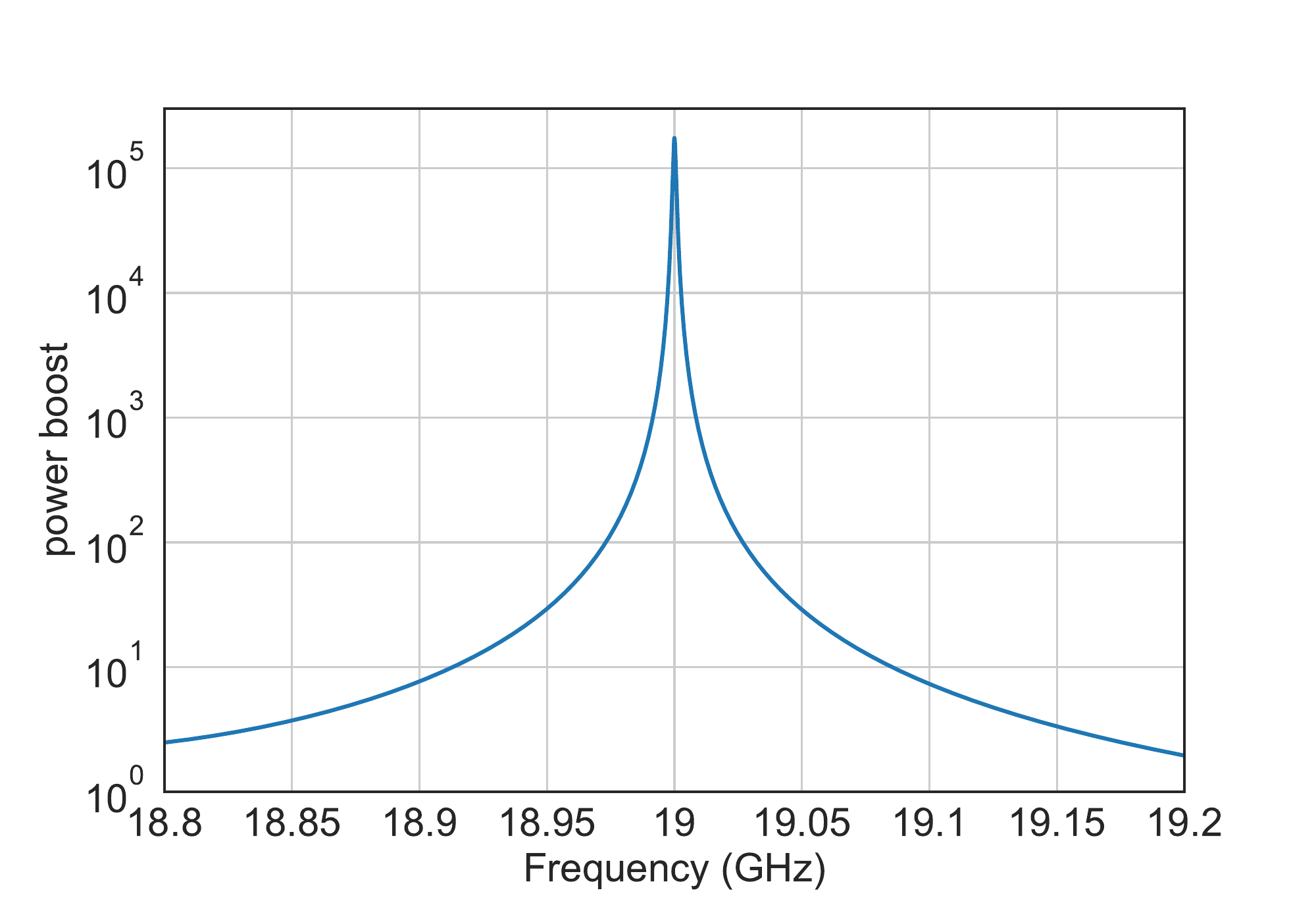}
\caption{\label{fig:power_boost} Calculated power boost from a simple T-RAX described in Sec.~\ref{sec:1D_DH}. The system consists of a short-circuited rectangular waveguide and a 1-mm thick sapphire slab. Table~\ref{tab:geometry} lists the exact dimensions.}
\end{figure}

Fig.~\ref{fig:power_boost} shows the calculated power boost from an example dielectric haloscope that consists of an 8-mm wide rectangular waveguide and a 1-mm thick sapphire slab. The 47.96-mm spacing between the mirror and the dielectric maximizes the power boost at $19\,$GHz. The peak power boost here is 175,000. 

So far, we have modeled a simple dielectric haloscope. While the model presented in this section is insightful and accurate for simple systems, it is not applicable for more complicated and powerful systems. They have internal reflections that modify wave propagation. Impedance mismatch boundaries, the sources of additional reflections, also source additional axion-induced traveling waves. The simple analytical model does not account for these contributions. In the next section, we study a more complex geometry using numerical simulations.

\section{Simulation}\label{sec:simulation}
In this section, we explain the detail of the T-RAX simulation. We describe the simulation geometry and the underlying reasoning behind it. We show reflectivity and axion signal power from simulation and discuss how to tune the system to scan a wider mass range.

\subsection{Geometry}
\begin{figure}
\centering 
\definecolor{wwzzqq}{rgb}{0.4,0.6,0.}
\definecolor{zzttqq}{rgb}{0.6,0.2,0.}
\begin{tikzpicture}[line cap=round,line join=round,>=triangle 45,x=1.0cm,y=1.0cm]
\clip(-0.5819991230340139,-0.297931374303697) rectangle (9.412422767436802,5.803262151258756);
\fill[line width=0.4pt,color=zzttqq,fill=zzttqq,fill opacity=0.25] (1.8011143755657386,4.) -- (1.8,0.) -- (2.,0.) -- (2.,4.) -- cycle;
\fill[line width=0.4pt,color=zzttqq,fill=zzttqq,fill opacity=0.25] (3.802409993480479,4.) -- (3.8,0.) -- (4.,0.) -- (4.,4.) -- cycle;
\fill[line width=0.4pt,color=zzttqq,fill=zzttqq,fill opacity=0.10000000149011612] (2.,4.) -- (1.8026726372819641,4.2) -- (1.6002785915256532,4.2) -- (1.8011143755657386,4.) -- cycle;
\fill[line width=0.4pt,color=zzttqq,fill=zzttqq,fill opacity=0.10000000149011612] (1.6002785915256532,4.2) -- (1.8011143755657386,4.) -- (1.8,0.) -- (1.6,0.2) -- cycle;
\fill[line width=0.4pt,color=zzttqq,fill=zzttqq,fill opacity=0.10000000149011612] (4.,4.) -- (3.803691331587368,4.196308668412632) -- (3.597043987852414,4.2) -- (3.802409993480479,4.) -- cycle;
\fill[line width=0.4pt,color=zzttqq,fill=zzttqq,fill opacity=0.10000000149011612] (3.597043987852414,4.2) -- (3.802409993480479,4.) -- (3.8,0.) -- (3.5976918554263198,0.1964999659922376) -- cycle;
\draw [line width=1.2pt] (0.,0.)-- (4.,0.);
\draw [line width=1.2pt] (0.,4.)-- (4.,4.);
\draw [line width=1.2pt] (0.,4.)-- (0.,0.);
\draw [line width=0.4pt,color=zzttqq] (1.8011143755657386,4.)-- (1.8,0.);
\draw [line width=0.4pt,color=zzttqq] (1.8,0.)-- (2.,0.);
\draw [line width=0.4pt,color=zzttqq] (2.,0.)-- (2.,4.);
\draw [line width=0.4pt,color=zzttqq] (2.,4.)-- (1.8011143755657386,4.);
\draw [line width=0.4pt,color=zzttqq] (3.802409993480479,4.)-- (3.8,0.);
\draw [line width=0.4pt,color=zzttqq] (3.8,0.)-- (4.,0.);
\draw [line width=0.4pt,color=zzttqq] (4.,0.)-- (4.,4.);
\draw [line width=0.4pt,color=zzttqq] (4.,4.)-- (3.802409993480479,4.);
\draw [line width=1.2pt] (4.,4.)-- (8.,2.2);
\draw [line width=1.2pt] (4.,0.)-- (8.00174842875182,1.798825013799662);
\draw [line width=1.2pt] (8.00174842875182,1.798825013799662)-- (8.5,1.8);
\draw [line width=1.2pt] (8.5,1.8)-- (8.5,2.2);
\draw [line width=1.2pt] (8.5,2.2)-- (8.,2.2);
\draw [line width=0.8pt] (8.,2.2)-- (8.00174842875182,1.798825013799662);
\draw (7.5108690984653474,1.162029262536949) node[anchor=north west] {standard};
\draw (4.987480343432139,2.2615057915157073) node[anchor=north west] {taper};
\draw [->,line width=0.4pt] (7.,3.5) -- (8.,3.5);
\draw [->,line width=0.4pt] (7.,3.5) -- (7.,4.5);
\draw [->,line width=0.4pt] (7.,3.5) -- (6.105329876457198,4.244201426547813);
\draw (7.898389514416876,3.3699944231910126) node[anchor=north west] {z};
\draw (7.042239758244895,4.640700903404167) node[anchor=north west] {y};
\draw (6.159053693983272,4.514531465652507) node[anchor=north west] {x};
\draw [line width=1.2pt,color=wwzzqq] (0.,4.2)-- (0.,4.6);
\draw [line width=1.2pt,color=wwzzqq] (0.5,4.2)-- (0.5,4.6);
\draw [line width=1.2pt,color=wwzzqq] (1.,4.2)-- (1.,4.6);
\draw [line width=1.2pt,color=wwzzqq] (1.5,4.2)-- (1.5,4.6);
\draw [line width=1.2pt,color=wwzzqq] (2.,4.2)-- (2.,4.6);
\draw [line width=1.2pt,color=wwzzqq] (2.5,4.2)-- (2.5,4.6);
\draw [line width=1.2pt,color=wwzzqq] (3.,4.2)-- (3.,4.6);
\draw [line width=1.2pt,color=wwzzqq] (3.5,4.2)-- (3.5,4.6);
\draw [line width=1.2pt,color=wwzzqq] (4.,4.2)-- (4.,4.6);
\draw [color=wwzzqq](1.6530023457096852,5.541911173058887) node[anchor=north west] {$B_e$};
\draw (7.519881201161895,0.7564846411923252) node[anchor=north west] {waveguide};
\draw (0.661671049089496,2.6129777966810477) node[anchor=north west] {rectangular};
\draw (0.7337678706618734,2.018179018708933) node[anchor=north west] {waveguide};
\draw [line width=1.2pt] (-0.2,0.2)-- (-0.2,4.2);
\draw [line width=1.2pt] (-0.2,4.2)-- (3.8,4.2);
\draw [line width=1.2pt] (-0.2,0.2)-- (0.,0.);
\draw [line width=1.2pt] (-0.2,4.2)-- (0.,4.);
\draw [line width=1.2pt] (3.8,4.2)-- (4.,4.);
\draw [line width=0.4pt,color=zzttqq] (2.,4.)-- (1.8026726372819641,4.2);
\draw [line width=0.4pt,color=zzttqq] (1.8026726372819641,4.2)-- (1.6002785915256532,4.2);
\draw [line width=0.4pt,color=zzttqq] (1.6002785915256532,4.2)-- (1.8011143755657386,4.);
\draw [line width=0.4pt,color=zzttqq] (1.8011143755657386,4.)-- (2.,4.);
\draw [line width=0.4pt,color=zzttqq] (1.6002785915256532,4.2)-- (1.8011143755657386,4.);
\draw [line width=0.4pt,color=zzttqq] (1.8011143755657386,4.)-- (1.8,0.);
\draw [line width=0.4pt,color=zzttqq] (1.8,0.)-- (1.6,0.2);
\draw [line width=0.4pt,color=zzttqq] (1.6,0.2)-- (1.6002785915256532,4.2);
\draw [line width=0.4pt,color=zzttqq] (4.,4.)-- (3.803691331587368,4.196308668412632);
\draw [line width=0.4pt,color=zzttqq] (3.803691331587368,4.196308668412632)-- (3.597043987852414,4.2);
\draw [line width=0.4pt,color=zzttqq] (3.597043987852414,4.2)-- (3.802409993480479,4.);
\draw [line width=0.4pt,color=zzttqq] (3.802409993480479,4.)-- (4.,4.);
\draw [line width=0.4pt,color=zzttqq] (3.597043987852414,4.2)-- (3.802409993480479,4.);
\draw [line width=0.4pt,color=zzttqq] (3.802409993480479,4.)-- (3.8,0.);
\draw [line width=0.4pt,color=zzttqq] (3.8,0.)-- (3.5976918554263198,0.1964999659922376);
\draw [line width=0.4pt,color=zzttqq] (3.5976918554263198,0.1964999659922376)-- (3.597043987852414,4.2);
\draw [line width=0.4pt] (-0.2,0.2)-- (1.6,0.2);
\draw [line width=0.4pt] (2.,0.203951631275401)-- (3.5976918554263193,0.19649996599223774);
\draw [line width=1.2pt] (8.5,2.2)-- (8.299692649102427,2.402192889808609);
\draw [line width=1.2pt] (8.299692649102427,2.402192889808609)-- (7.800431075130478,2.402192889808609);
\draw [line width=1.2pt] (7.800431075130478,2.402192889808609)-- (8.,2.2);
\draw [line width=1.2pt] (7.800431075130478,2.402192889808609)-- (3.8,4.2);
\draw [line width=0.4pt] (8.00174842875182,1.7988250137996624)-- (7.805975493408163,1.9934337907015833);
\draw [line width=0.4pt] (7.805975493408163,1.9934337907015833)-- (7.800431075130492,2.4021928898086062);
\draw [line width=0.4pt] (7.805975493408163,1.9934337907015833)-- (4.,0.26039583081668594);
\draw [line width=0.8pt] (0.1,1.)-- (1.7,1.);
\draw [line width=0.8pt] (0.8842288202810422,1.0405544621344622) -- (0.8842288202810422,0.9594455378655375);
\draw [line width=0.8pt] (0.9157711797189573,1.0405544621344622) -- (0.9157711797189573,0.9594455378655375);
\draw [line width=0.8pt] (2.098179440046084,1.)-- (3.698179440046084,1.);
\draw [line width=0.8pt] (2.8824082603271264,1.0405544621344622) -- (2.8824082603271264,0.9594455378655375);
\draw [line width=0.8pt] (2.9139506197650413,1.0405544621344622) -- (2.9139506197650413,0.9594455378655375);
\begin{scriptsize}
\draw [fill=wwzzqq,shift={(0.,4.6)}] (0,0) ++(0 pt,2.25pt) -- ++(1.9485571585149868pt,-3.375pt)--++(-3.8971143170299736pt,0 pt) -- ++(1.9485571585149868pt,3.375pt);
\draw [fill=wwzzqq,shift={(0.5,4.6)}] (0,0) ++(0 pt,2.25pt) -- ++(1.9485571585149868pt,-3.375pt)--++(-3.8971143170299736pt,0 pt) -- ++(1.9485571585149868pt,3.375pt);
\draw [fill=wwzzqq,shift={(1.,4.6)}] (0,0) ++(0 pt,2.25pt) -- ++(1.9485571585149868pt,-3.375pt)--++(-3.8971143170299736pt,0 pt) -- ++(1.9485571585149868pt,3.375pt);
\draw [fill=wwzzqq,shift={(1.5,4.6)}] (0,0) ++(0 pt,2.25pt) -- ++(1.9485571585149868pt,-3.375pt)--++(-3.8971143170299736pt,0 pt) -- ++(1.9485571585149868pt,3.375pt);
\draw [fill=wwzzqq,shift={(2.,4.6)}] (0,0) ++(0 pt,2.25pt) -- ++(1.9485571585149868pt,-3.375pt)--++(-3.8971143170299736pt,0 pt) -- ++(1.9485571585149868pt,3.375pt);
\draw [fill=wwzzqq,shift={(2.5,4.6)}] (0,0) ++(0 pt,2.25pt) -- ++(1.9485571585149868pt,-3.375pt)--++(-3.8971143170299736pt,0 pt) -- ++(1.9485571585149868pt,3.375pt);
\draw [fill=wwzzqq,shift={(3.,4.6)}] (0,0) ++(0 pt,2.25pt) -- ++(1.9485571585149868pt,-3.375pt)--++(-3.8971143170299736pt,0 pt) -- ++(1.9485571585149868pt,3.375pt);
\draw [fill=wwzzqq,shift={(3.5,4.6)}] (0,0) ++(0 pt,2.25pt) -- ++(1.9485571585149868pt,-3.375pt)--++(-3.8971143170299736pt,0 pt) -- ++(1.9485571585149868pt,3.375pt);
\draw [fill=wwzzqq,shift={(4.,4.6)}] (0,0) ++(0 pt,2.25pt) -- ++(1.9485571585149868pt,-3.375pt)--++(-3.8971143170299736pt,0 pt) -- ++(1.9485571585149868pt,3.375pt);
\draw [fill=black,shift={(0.1,1.)},rotate=90] (0,0) ++(0 pt,2.25pt) -- ++(1.9485571585149868pt,-3.375pt)--++(-3.8971143170299736pt,0 pt) -- ++(1.9485571585149868pt,3.375pt);
\draw [fill=black,shift={(1.7,1.)},rotate=270] (0,0) ++(0 pt,2.25pt) -- ++(1.9485571585149868pt,-3.375pt)--++(-3.8971143170299736pt,0 pt) -- ++(1.9485571585149868pt,3.375pt);
\draw [fill=black,shift={(2.098179440046084,1.)},rotate=90] (0,0) ++(0 pt,2.25pt) -- ++(1.9485571585149868pt,-3.375pt)--++(-3.8971143170299736pt,0 pt) -- ++(1.9485571585149868pt,3.375pt);
\draw [fill=black,shift={(3.698179440046084,1.)},rotate=270] (0,0) ++(0 pt,2.25pt) -- ++(1.9485571585149868pt,-3.375pt)--++(-3.8971143170299736pt,0 pt) -- ++(1.9485571585149868pt,3.375pt);
\end{scriptsize}
\end{tikzpicture}
\caption{\label{fig:schematicGeometry} Schematic T-RAX geometry simulated in this paper, including the dielectrics (brown color). This image is not to scale. Table~\ref{tab:geometry} lists the exact simulation dimensions.}
\end{figure}
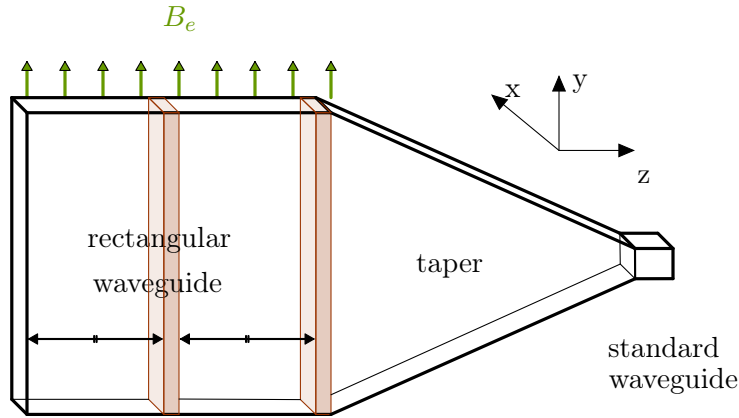

Fig.~\ref{fig:schematicGeometry} shows the simulation geometry. A long, thin rectangular waveguide forms the detector body. Its two vertical walls are at a distance slightly larger than the free-space half-wavelength, $k_xa > \pi$, of our 19\,GHz design frequency. A conductive mirror shorts the waveguide on the left side. Two 1-mm thick equally-spaced dielectric slabs divide the waveguide volume along the $z$ axis. Dielectrics can longitudinally reflect the traveling waves and further enhance the signal power. We choose C-cut sapphires for their high relative permittivity, low loss, mechanical strength, and availability. Our simulations show that anisotropy of the sapphire's permittivity hardly affects the result. A hexahedrally-tapered waveguide connects the detector section to a standard waveguide (WR42 in our case). 

\subsection{Dielectric positioning}\label{sec:dielectric_position}
\begin{figure}
\centering 
\includegraphics[width=0.6\textwidth]{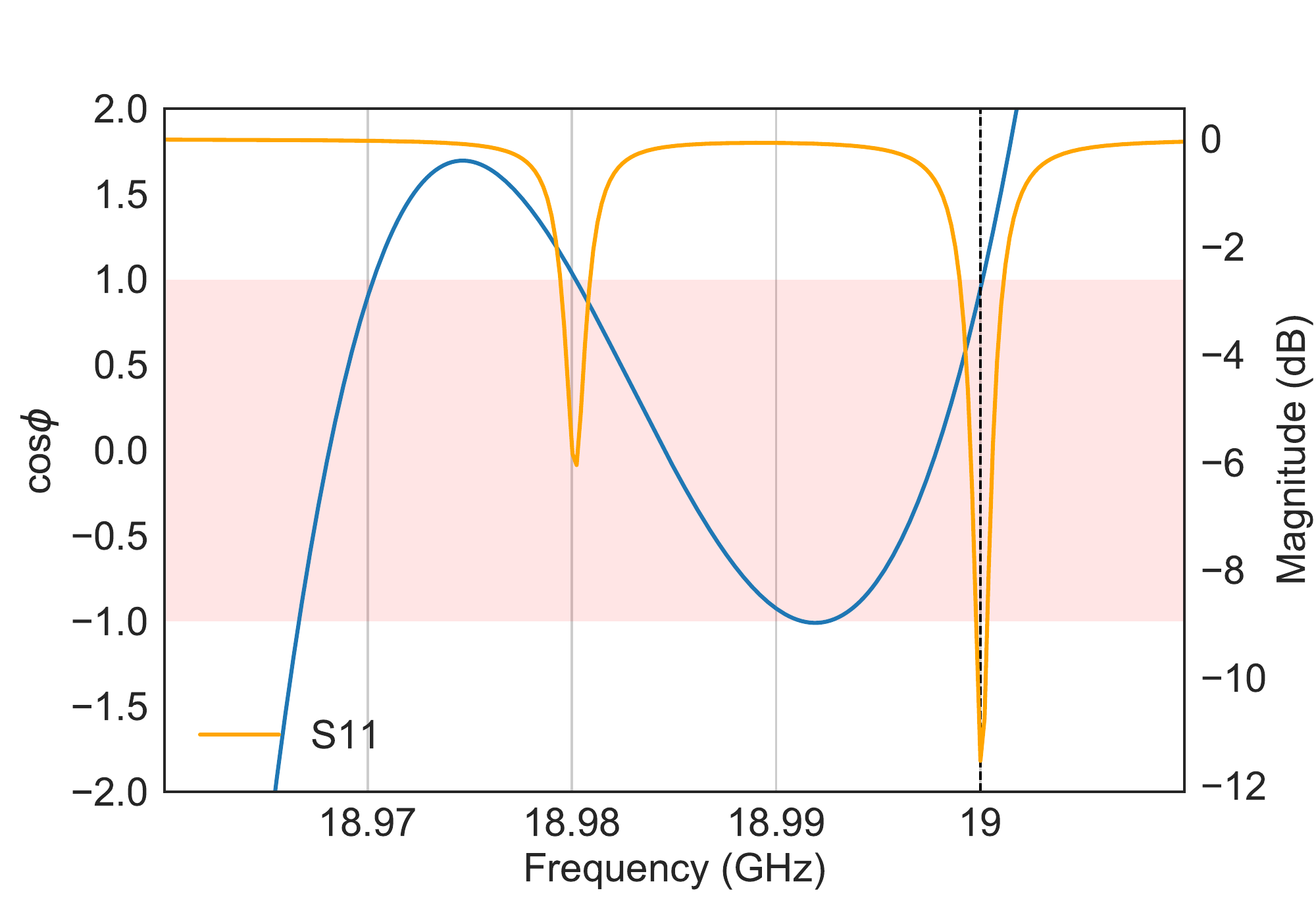}
\caption{\label{fig:dispersion_s11} Dispersion and reflection magnitude of the two dielectric structure. The $y-$axis on the left shows the $\cos{\phi}$, where $\phi$ is the phase difference $\phi$ between the input and the reflected waves off two dielectric T-RAX structure. The resonances, or the reflectivity dips, occur at $\cos{\phi}=1$. The horizontal pink band represents the passband. Check Appendix~\ref{ch:dispersion} for the calculation details.}
\end{figure}

The $z$ position of the dielectrics determines the resonance frequency. The system resonates when the reflectivity is real, i.e., when the phase difference $\phi$ between the input and the reflected wave is zero, or $\cos{\phi}=1$. In general, the signal power is maximal when a T-RAX forms a parallel resonator\footnote{The simple model in Sec.~\ref{sec:theory} does not consider the dielectric thickness.}. For a fixed dielectric permittivity and thickness, we can find the resonance using the dispersion calculation in Appendix~\ref{ch:dispersion}. The two dielectric-separated sections form two coupled resonators that have two resonance modes: anti-symmetric and symmetric. Only the latter couples effectively with the axion-induced electric field. Fig.~\ref{fig:dispersion_s11} shows a clear correlation between dispersion and reflectivity for the two-dielectric system. 

The dispersion also guides how to calculate the spacing to boost the desired frequency. Given the waveguide geometry, dielectric thickness, and dielectric's relative permittivity, one can find spacings that satisfy the standing wave condition, \emph{i.e.} $\cos{\phi}=1$. For example, 49.108\,mm vacuum spacings between the dielectrics can enhance the signal power at 19\,GHz. We can further optimize the system, for example, by having different gap distances, using microwave filter theory as in the case of Ref.~\cite{Melc_n_2018}. However, for practical purposes, we vary the spacing for tuning rather than optimization.

\subsection{Taper design}
The taper transforms the TE10 mode of the resonator to that of the standard waveguide. We want to minimize reflection and mode conversion in the taper. In our example, the height (the E-plane or the $y$ axis) decreases from 100 to 4.318\,mm of WR42. The width (the H-plane or the $x$ axis) increases from 8 to 10.668\,mm of WR42. The taper barely converts the TE10 mode signal to TE/TM $0m$ high order modes because the overlap among them is small. However, a realistic taper might generate unwanted higher-order modes, and they need to be carefully monitored in future R\&D.

Reflections and mode conversions at impedance mismatches distort the booster behavior and should be avoided. The H-plane taper changes the TE10-mode wave impedance. In our example, the resonator's wave impedance is four times higher than inside the standard WR42 waveguide. The E-plane taper also adds a reactance whose magnitude depends on the propagation constant, taper angle, and waveguide height~\cite{1129671}. Not surprisingly, a longer taper with a lower launch angle generally reflects less. We empirically find that the taper height twice its base length insignificantly distorts the power boost peaks. We plan to optimize the taper using nonlinear profiles~\cite{ParabolicTaper} or a dielectric lens in the future.

\begin{table}[tbp]
\centering
\begin{tabular}{|c|c|}
\hline
design frequency & 19\,GHz \\
TE10 mode cutoff frequency & 18.737\,GHz \\
\hline
dielectric& sapphire\\
\hline
relative permittivity & 9.4\\
thickness   & 1\,mm\\
loss $\tan{\delta}$ & $10^{-5}$ \\
\hline
& $6 \times 10^{7}$\,S/m\\
waveguide wall conductivity & $2 \times 10^{8}$\,S/m\\
& $1 \times 10^{9}$\,S/m\\
\hline
geometry & \\
\hline
rectangular waveguide width ($a$, $x$ axis) & 8\,mm \\
rectangular waveguide height ($b$, $y$ axis) & 100\,mm \\
distance between dielectrics & 49.108\,mm\\
taper length & 300\,mm \\
WR42 waveguide width & 10.668\,mm \\
WR42 waveguide height & 4.318\,mm \\
\hline
\end{tabular}
\caption{\label{tab:geometry} Key quantities of simulations}
\end{table}

\subsection{Simulation techniques}
We simulate the reflectivity and axion-induced power of the proposed geometry using COMSOL Multiphysics\textsuperscript{\textregistered} and its RF module. The 3D simulation geometry is cut symmetrically along the $x$ and $y$ axes to save simulation resources. The reference plane for the reflectivity and the signal power is the WR42 waveguide's TE10-mode port. The simulations include resistive loss of copper and dielectric loss inside the sapphire. The latter is subdominant because the field is weak around the dielectrics. To simulate the system's response under axion-induced electric field, we distribute the axion-induced current $J_a = \frac{\alpha}{2 \pi}C_{a\gamma} B_e \dot{\theta}$ vertically, parallel to the $y$-axis. The current decays slowly along the $z$ axis to suppress the $\nabla \theta$ contribution. We integrate the time-averaged Poynting vector normal to the waveguide port. The power boost is the ratio of this integral to the axion-induced power $P_0$ of the waveguide's cross-sectional area. 



\subsection{Simulation results}
\begin{figure}
\centering 
\includegraphics[width=.8\textwidth]{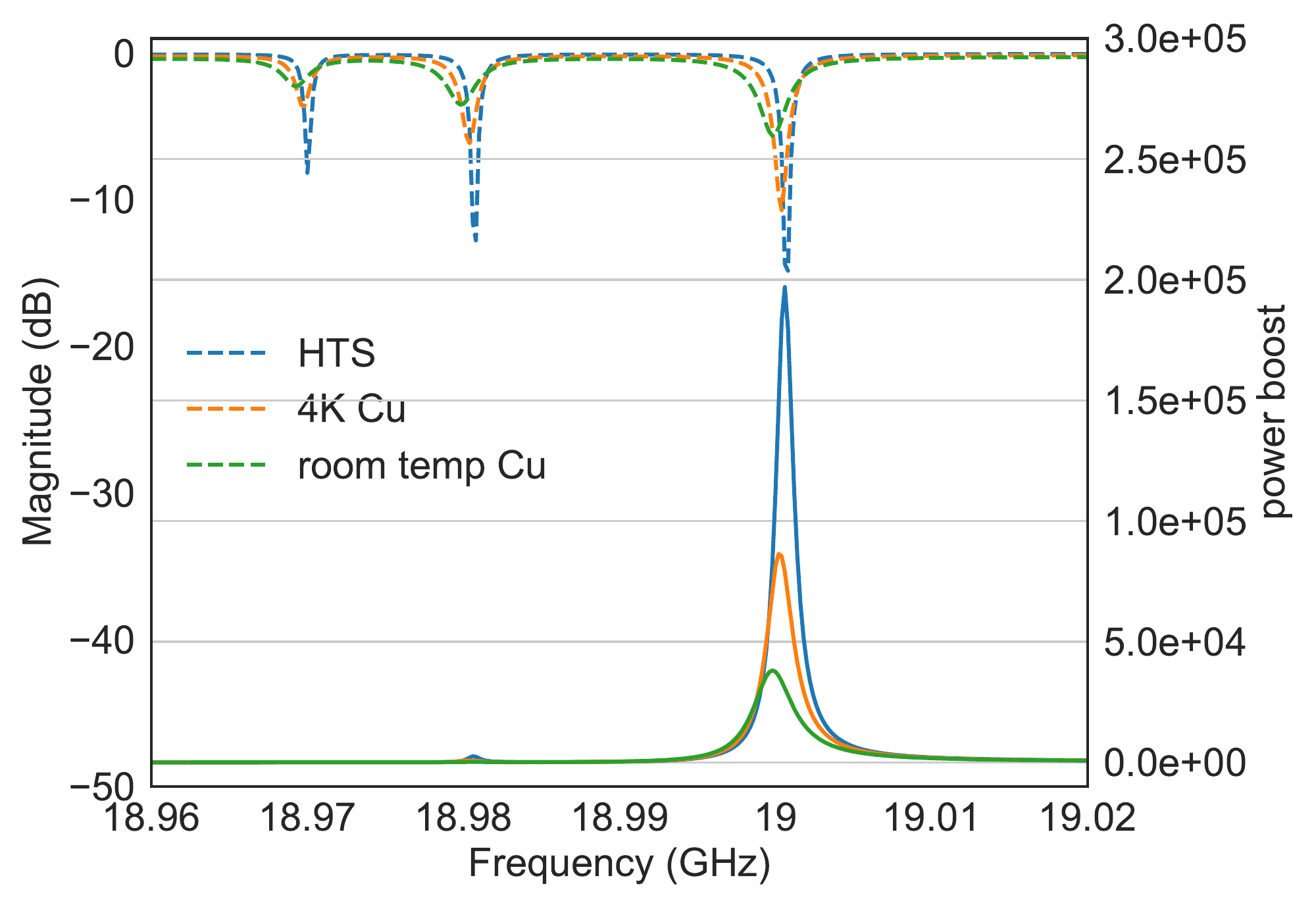}
\caption{\label{fig:noTaperSim} Reflectivity magnitude (dashed) and axion-induced power (solid) from T-RAX simulation. Different colors represent the varying surfaces, simulated with conductivities of $1 \times 10^{9}$, $2 \times 10^{8}$, and $6 \times 10^{7}$\,[S/m] for high-temperature superconductor (HTS), 4K copper, and room temperature copper.}
\end{figure}

\begin{figure}
\centering 
\includegraphics[width=.6\textwidth]{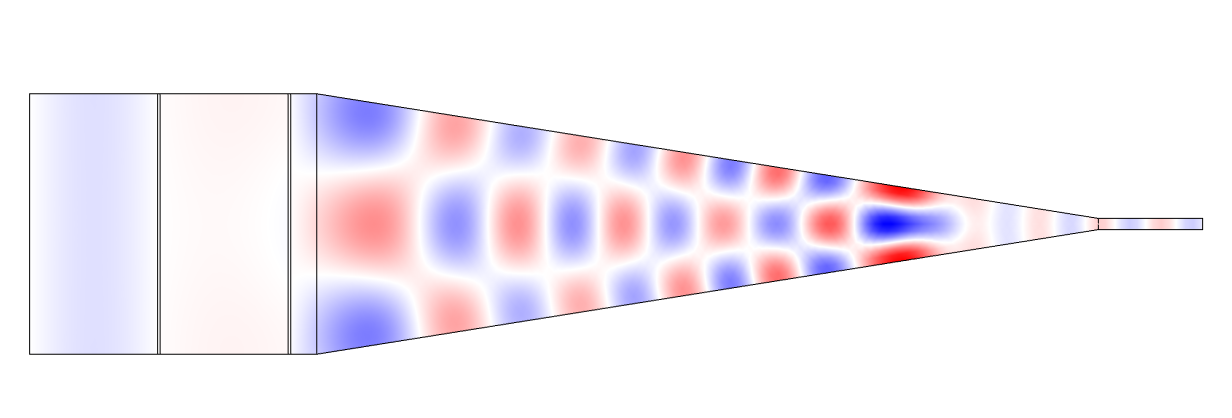}
\includegraphics[width=.6\textwidth]{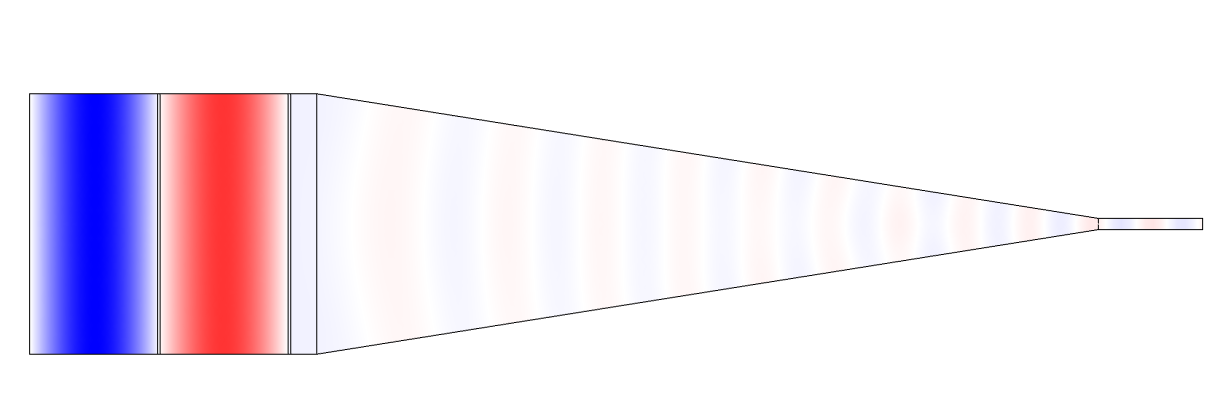}
\includegraphics[width=.6\textwidth]{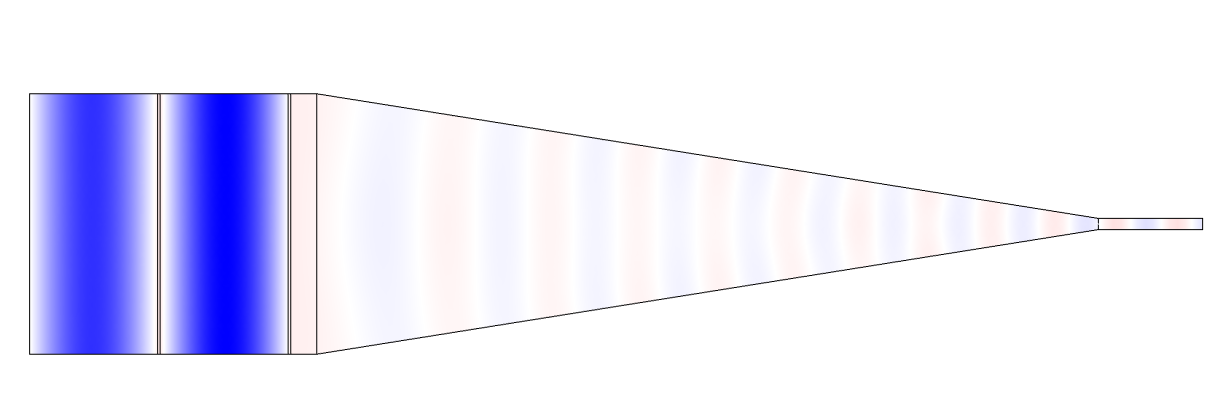}
\caption{\label{fig:noTaperField} Vertical electric field inside the T-RAX at taper-conversion (top, 18.97\,GHz), asymmetric-resonance (middle, 18.98\,GHz), and symmetric resonance (bottom, 19.00\,GHz) modes. The blue and red color indicate opposite directions.}
\end{figure}

Fig.~\ref{fig:noTaperSim} shows the simulation results. The reflectivity (dashed lines) and the power boost (solid lines) peak at 18.97, 18.98, and 19\,GHz. Fig.~\ref{fig:noTaperField} shows the vertical field map at the peaks: these frequencies indeed correspond to the mode-conversion inside the taper, asymmetric, and symmetric resonances of the detector. The axion-induced power is maximum for the latter case when both resonators oscillate in phase.

The peak signal power largely depends on the resistive loss of copper. Temperature, surface roughness, and machining precision all affect the reflectivity and the power boost. We simulate three different surface conductivities to represent the different surface losses: $6 \times 10^{7}$\,[S/m] for room temperature, $2 \times 10^{8}$\,[S/m] for 4\,K, and $1 \times 10^{9}$\,[S/m] for a high-temperature superconductor (HTS) coated system. Higher conductivity slightly shifts the peak and makes it narrower. The peak power boost for $2 \times 10^{8}$\,[S/m] is about 86,000 at 19\,GHz. The boost peak is about 1\,MHz wide.

The peak boost power drops at a higher frequency as the sheet resistance $R_s$ increases. We simulate the T-RAX system in a wider frequency range by scaling the default geometry. The peak power boost decreases by about $\nu^{-0.64}$, close to the expectation of $\nu^{-\frac{2}{3}}$ from the anomalous skin effect.

\subsection{Tuning}
As already discussed in Sec.~\ref{sec:dielectric_position}, the resonant frequency critically depends on the dielectric position. Moving the dielectric along the waveguide length ($z$ axis) is the most practical way to tune T-RAX because we can not easily change the dielectric's permittivity and thickness. The theoretical low-frequency limit is the cutoff frequency. However, the dielectric spacing (roughly a half guided-wavelength) rapidly diverges to infinity near the cutoff frequency, and such a system would not fit inside a real magnet\footnote{Our simulation shows that the elongated waveguide section can also be warped to reduce the overall profile and fit inside a small solenoid magnet. Such geometry, unfortunately, is even harder to realize, and we do not propose it as the default design.}. In other words, the available magnet space limits the low-frequency bound. 

If we reduce the dielectric spacing, the resonant frequency will increase, and the transverse resonance and the $P_\text{sig}$ will gradually decrease. The decreasing $P_\text{sig}$ sets the high-frequency limit. A simple model relates the peak boost frequency and the tuning range. From Eq.~\ref{eq:e_tr_transverse},
\begin{equation}
    \frac{{E^{tr}_y}^2}{E_0^2} = \frac{1}{\cos^2{\frac{k_x a}{2}}} = \frac{2}{1+\cos{k_x a}} \simeq \frac{4}{(k_x a - \pi)^2}.
\end{equation}
The last approximation is valid near the cutoff; $k_x a \simeq \pi$. The peak power boost occurs at $k_p$ due to the geometry and the conductor loss. We want to find $k_d$ slightly higher than $k_p$, where the signal power or ${E^{tr}_y}^2$ drops by a fraction $p$: 
\begin{equation}
    p \frac{4}{(k_p a - \pi)^2} = \frac{4}{(k_d a - \pi)^2},
\end{equation}
or
\begin{equation}
    k_d a = \sqrt{\frac{1}{p}} (k_p a - \pi) + \pi.
\end{equation}
The tuning range $\Delta k$ is the difference between $k_d$ and $k_p$:
\begin{equation}
    \Delta k = k_p - k_d = (\sqrt{\frac{1}{p}} - 1) \frac{k_p a - \pi}{a}.
\end{equation}
Notice that $\Delta k$ decreases as we operate closer to the cutoff frequency. Operation near the cutoff frequency significantly increases $P_\text{sig}$ at the price of reduced bandwidth.

\begin{figure}
\centering 
\includegraphics[width=.6\textwidth]{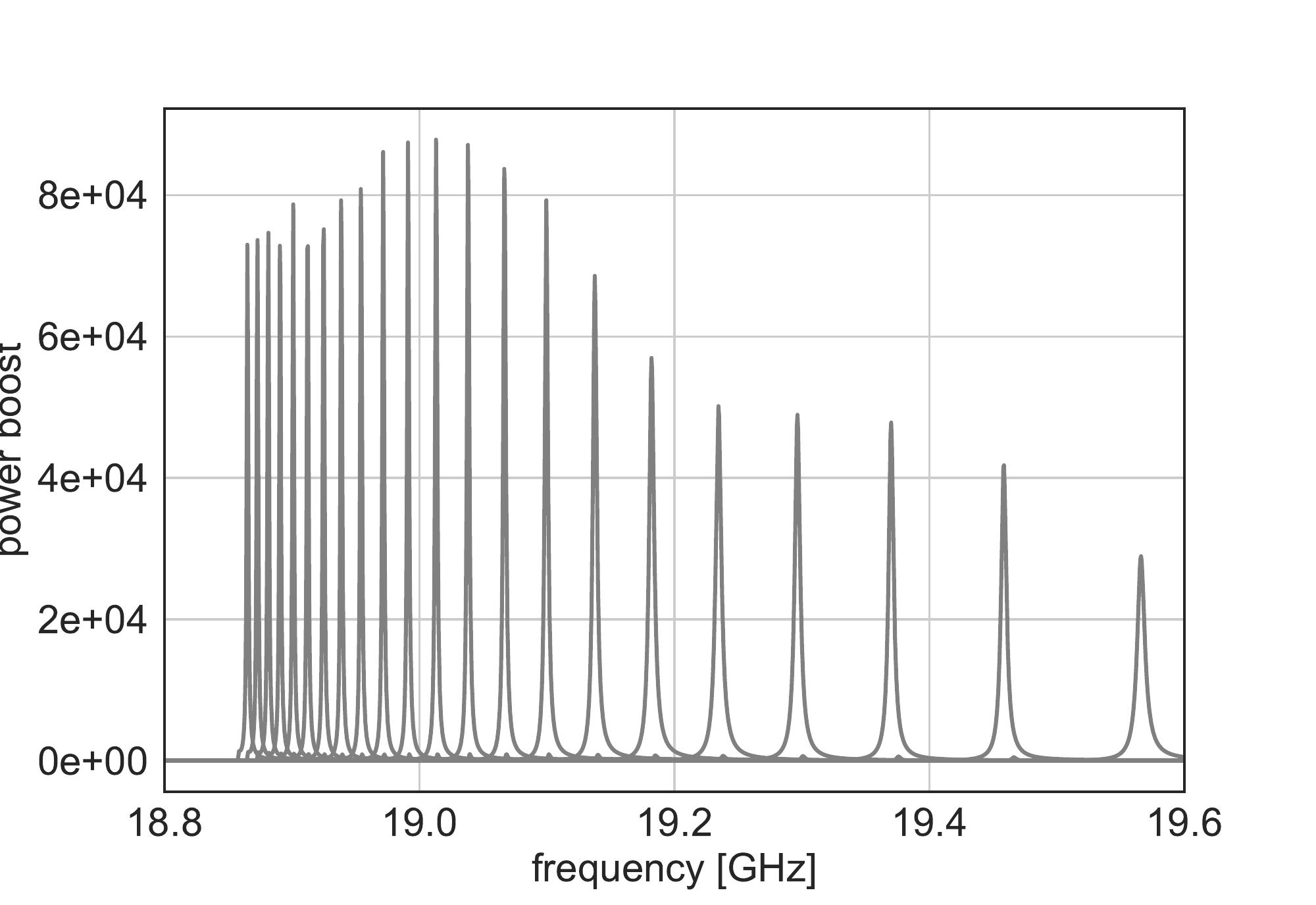}
\caption{\label{fig:tuning} T-RAX can be tuned over 300\,MHz by moving the dielectrics. The spacing between the dielectrics changes from 28 to 70\,mm in 2\,mm steps. The cutoff frequency is 18.737\,GHz.}
\end{figure}

Numerical simulation with loss provides a more realistic tuning scenario. Fig.~\ref{fig:tuning} shows how the boost frequency shifts if we move the dielectrics. The peak power boost is about 90,000, and it drops to half around 19.4\,GHz. The lower limit is 18.86\,GHz, with 70-mm spacing between the dielectrics. Again, this limit will decrease with larger spacing, but the larger detector hardly fits inside a real magnet. We claim a practical tuning range of 500\,MHz, and the relative tuning range $\frac{\delta \nu}{\nu} \approx$ 2.5\,\%. The dielectric has to be placed with a $50\,\mu$m precision. Such control is readily feasible with existing cryogenic stepper motors.


\begin{figure}
\begin{center}
\definecolor{zzttqq}{rgb}{0.6,0.2,0.}
\begin{tikzpicture}[line cap=round,line join=round,>=triangle 45,x=1.0cm,y=1.0cm]
\clip(-2.5330698952034294,-0.16462597078115154) rectangle (3.712409453679059,7.151506980766965);
\fill[line width=0.pt,color=zzttqq,fill=zzttqq,fill opacity=0.10000000149011612] (0.6,2.8) -- (0.6,0.) -- (2.,0.) -- (2.,2.8) -- cycle;
\fill[line width=2.pt,color=zzttqq,fill=zzttqq,fill opacity=0.10000000149011612] (2.,2.8) -- (1.7986745906424062,2.963183433513259) -- (0.4009333006726468,2.963183433513259) -- (0.6,2.8) -- cycle;
\fill[line width=0.pt,color=zzttqq,fill=zzttqq,fill opacity=0.10000000149011612] (0.39642445780177665,0.1631920107028738) -- (0.4009333006726468,2.963183433513259) -- (0.6,2.8) -- (0.6,0.) -- cycle;
\draw [line width=1.2pt] (0.,4.)-- (0.,0.);
\draw [line width=1.2pt] (0.,4.)-- (1.,4.);
\draw [line width=1.2pt] (1.,4.)-- (1.,3.);
\draw [line width=1.2pt] (1.,3.)-- (0.2,3.);
\draw [line width=1.2pt] (0.2,3.)-- (0.2,0.);
\draw [line width=1.2pt] (1.6,4.)-- (1.6,3.);
\draw [line width=1.2pt] (1.6,3.)-- (2.4,3.);
\draw [line width=1.2pt] (2.4,3.)-- (2.4,0.);
\draw [line width=1.2pt] (1.6,4.)-- (2.6,4.);
\draw [line width=1.2pt] (2.6,4.)-- (2.6,0.);
\draw [line width=1.2pt,color=zzttqq] (2.,0.)-- (2.,2.8);
\draw [line width=1.2pt] (2.6,4.)-- (0.4,6.);
\draw [line width=1.2pt] (1.6,4.)-- (-0.6,6.);
\draw [line width=1.2pt] (1.,4.)-- (-1.2,6.);
\draw [line width=1.2pt] (0.,4.)-- (-2.2,6.02);
\draw [line width=1.2pt,color=zzttqq] (2.,2.8)-- (1.7986745906424062,2.963183433513259);
\draw [line width=1.2pt,color=zzttqq] (1.7986745906424062,2.963183433513259)-- (0.4009333006726468,2.963183433513259);
\draw [line width=1.2pt,color=zzttqq] (0.4009333006726468,2.963183433513259)-- (0.6,2.8);
\draw [line width=1.2pt,color=zzttqq] (0.6,2.8)-- (2.,2.8);
\draw [line width=1.2pt,color=zzttqq] (0.39642445780177665,0.1631920107028738)-- (0.4009333006726468,2.963183433513259);
\draw [line width=1.2pt,color=zzttqq] (0.6,2.8)-- (0.6,0.);
\draw [shift={(1.2491265882379228,2.87996249605615)},line width=1.2pt]  plot[domain=3.0685520882866717:6.210144741876465,variable=\t]({1.*0.05122454094069229*cos(\t r)+0.*0.05122454094069229*sin(\t r)},{0.*0.05122454094069229*cos(\t r)+1.*0.05122454094069229*sin(\t r)});
\draw [line width=1.2pt] (1.300214550393572,2.876224352483785)-- (1.2971269841301953,4.161488668226302);
\draw [line width=1.2pt] (1.1980386260822735,2.8837006396285143)-- (1.1973638477701989,4.161488668226302);
\draw [line width=1.2pt] (1.2489654700253694,4.1597686141511305) circle (0.04819221956800561cm);
\draw [line width=1.2pt] (1.6,3.)-- (1.2993194714705363,3.248819756571713);
\draw [line width=1.2pt] (1.1978046200087065,3.3268241871295214)-- (1.,3.503189614949408);
\draw (-1.9680027160188232,3.166296348622869) node[anchor=north west] {E-plane};
\draw (-2.2554930352530964,6.814449365112987) node[anchor=north west] {H-plane};
\draw (1.4224003590888132,6.041199540965625) node[anchor=north west] {tuning rod};
\draw [->,line width=0.8pt] (1.769371434026729,5.4463919839291925) -- (1.3196905245598898,4.254826795854646);
\draw [->,line width=1.2pt] (0.9168139356078499,4.375738381263615) -- (-0.32236847488470743,5.4860458210649545);
\end{tikzpicture}
\caption{\label{fig:tuning_implementation} Cross-sectional view of T-RAX with its dielectric. The tuning rod enters the system from the tuning slit on the H-plane, moves the dielectric (brown), and changes the resonant frequency. The gaps between the dielectric and the E-plane walls minimally disturb the signal power.}
\end{center}
\end{figure}
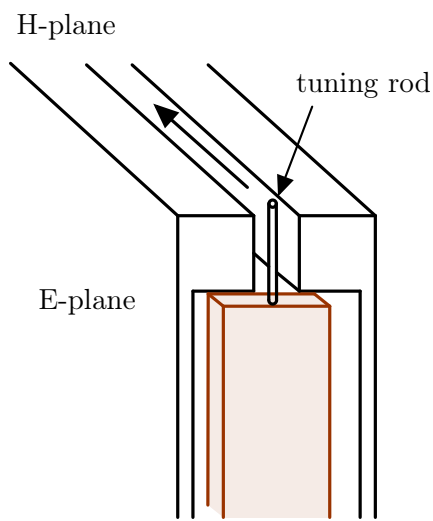

Dielectric displacement is easy to implement and control. We can move the dielectric via metallic rods. The rods can enter the system i) from the top, as shown in Fig.~\ref{fig:tuning_implementation}, through slits on the waveguide's H-planes, or ii) from the back through slits on the taper. The first option is easier to implement and can be used to move the mirror-side dielectric. TE11 mode leaks insignificantly through the slits. The second option is harder to implement, but the field in the taper is far weaker and poses less trouble. Dielectric displacement requires only two actuators on the top and bottom of the dielectric and does not deform or change the main metallic waveguide structure.

To scan a broader range, we need to vary the waveguide width. Ref~\cite{Kuo_2020} already suggested using \emph{tuning slits} along with quarter-wave deep corrugations to tune the \emph{thin shell cavity}. T-RAX can use the same slits on its waveguide and taper. As an alternative to the corrugations, flexible conducting materials such as bellows for foil filling the slit can also be used to contain the energy loss via surface current. The irregular surface of the foil does not significantly modify the electromagnetic behavior because TE10 mode resonance relies mainly on the E-plane walls and less on the H-plane walls or the mirror at the end. Surprisingly, we don't need to change the dielectrics according to the varying width. Our simulations show that gaps between the dielectric and the E-plane walls do not significantly degrade the power boost, as long as the dielectric is covering $\sim10$\% of the center of the waveguide. We observe only a slight shift of the peak power boost frequency due to the gap. By increasing the 8\,mm waveguide width by 1\,mm, we can easily achieve a tuning range $>10$\,\%. Only the dielectric thickness limits the tuning range. The waveguide width has to be controlled with a precision of $\sim \lambda/2Q$, or $1\,\mu$m-level at 20\,GHz. Such control is challenging yet achieved in space telescopes~\cite{Warden2012CryogenicNF}.



\section{Solenoid magnet application}\label{sec:packaging}
A smaller axion dark matter detector offers several advantages. A solenoid magnet is usually stronger than its dipole counterparts and is more readily available from commercial suppliers. Even among the solenoid magnets, smaller bore ones tend to have a stronger field~\cite{https://doi.org/10.48550/arxiv.2203.14923}. Unfortunately, T-RAX in Fig.~\ref{fig:schematicGeometry} is still relatively long ($\sim400\,$mm) and can fit only in large solenoids.

\begin{figure}
\centering 
\includegraphics[width=.4\textwidth]{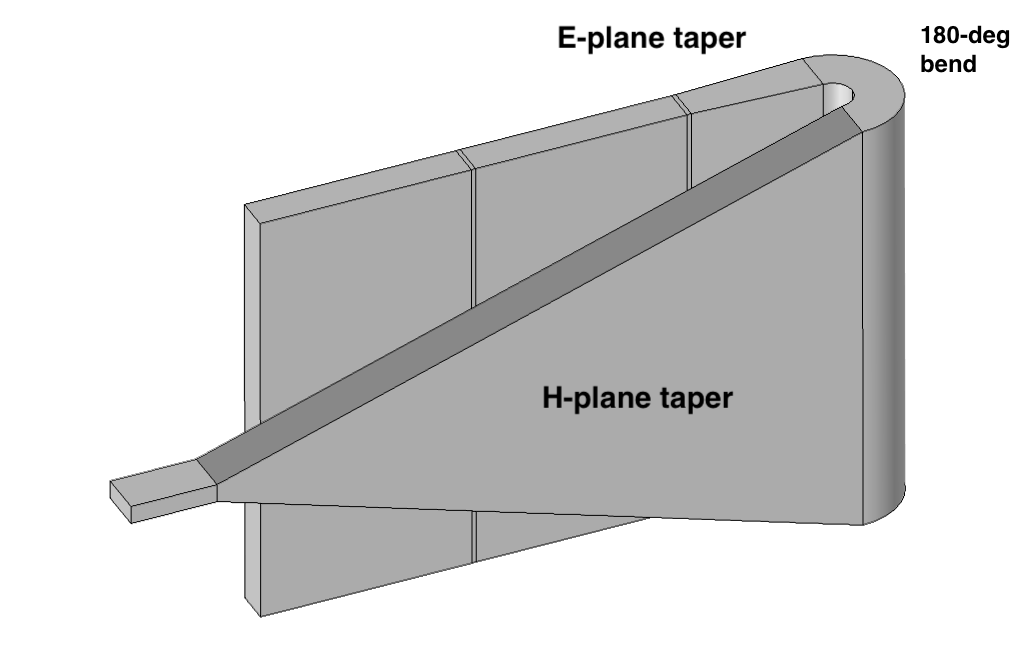}
\includegraphics[width=.5\textwidth]{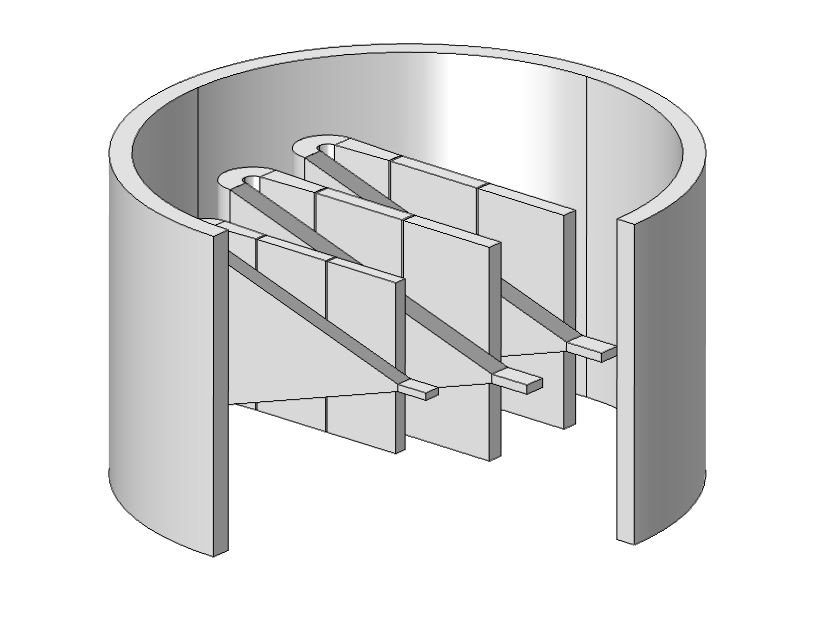}
\caption{\label{fig:packaging} A T-RAX with a folded taper (left). An array of T-RAX with folded tapers installed inside a solenoid magnet (right). The overall profile is 200\,mm.}
\end{figure}

We propose to \emph {fold} the taper to fit a fixed-width T-RAX inside a small conventional solenoid magnet, as shown in the left picture of Fig.~\ref{fig:packaging}. The right picture of Fig.~\ref{fig:packaging} shows how a T-RAX array can simultaneously operate inside a magnet to combine the signal power or scan different mass ranges. The folding has to reflect or convert the modes minimally. The larger guided wavelength near the cutoff frequency is particularly problematic~\cite{10.5555/27077}. An E-plane taper can expand to a slightly larger width, e.g., 8.5\,mm over the 30\,mm length. A $180^\circ$ bend with a 10-mm bend radius follows the E-plane taper. These components reflect -16.5 and -25\,dB\footnote{Although the $180^\circ$ bend appears to be very reflective, many commercial waveguide H-bends guarantee a return loss above 30 dB with an even shorter bend radius~\cite{wg_bend}.}, which are subdominant to the reflection at the input of a typical amplifier. The overall profile is about 140\,mm, which fits inside a proposed 30-T 160-mm-bore solenoid magnet~\cite{10.1007/978-3-030-43761-9_2, https://doi.org/10.48550/arxiv.2203.14923}. The taper after the bend tolerates more deformation and can be made of a flexible waveguide for easier packaging. The taper inside the strong magnetic field can source axion-induced emission and distort the signal. Fortunately, the transverse resonance inside the E-plane taper is small, and the signal power is minimally affected. 

\section{Projected sensitivity}\label{sec:sensitivity}
At the 19\,GHz design frequency, T-RAX's sensitivity to the axion-photon coupling $C_{a\gamma}$ is
\begin{equation}
\begin{split}
    C_{a\gamma} = 15.1 ~&\Big(\frac{300 \text{\,MeV} / \text{cm}^3}{\rho_a}\Big)^\frac{1}{2} \Big(\frac{80,000}{\beta^2}\Big)^\frac{1}{2}  \Big(\frac{8\,\text{mm} \times 100\,\text{mm}}{A}\Big)^\frac{1}{2} 
    \Big(\frac{10 \text{\,T}}{B_e}\Big)\\
    &\times\Big(\frac{T_\text{sys}}{0.9\,\text{K}}\Big)^\frac{1}{2} \Big(\frac{SNR}{5}\Big)^\frac{1}{2} \Big(\frac{0.85}{\eta}\Big)^\frac{1}{2} \Big(\frac{\Delta\nu_a}{19 ~\text{kHz}}\Big)^\frac{1}{4} \Big(\frac{1~\text{day}}{\tau}\Big)^\frac{1}{4}.
\end{split}
\label{eq:sensitivity}
\end{equation}
Here, $\rho_a$ is the local dark matter density, $A$ is the  waveguide's cross-sectional area, $T_{sys}$ is the system noise temperature, $SNR$ is the signal to noise ratio, $\eta$ is the efficiency, $\Delta\nu_a$ is the axion line width, and $\tau$ is the integration time.

\begin{figure}
\centering 
\includegraphics[width=.8\textwidth]{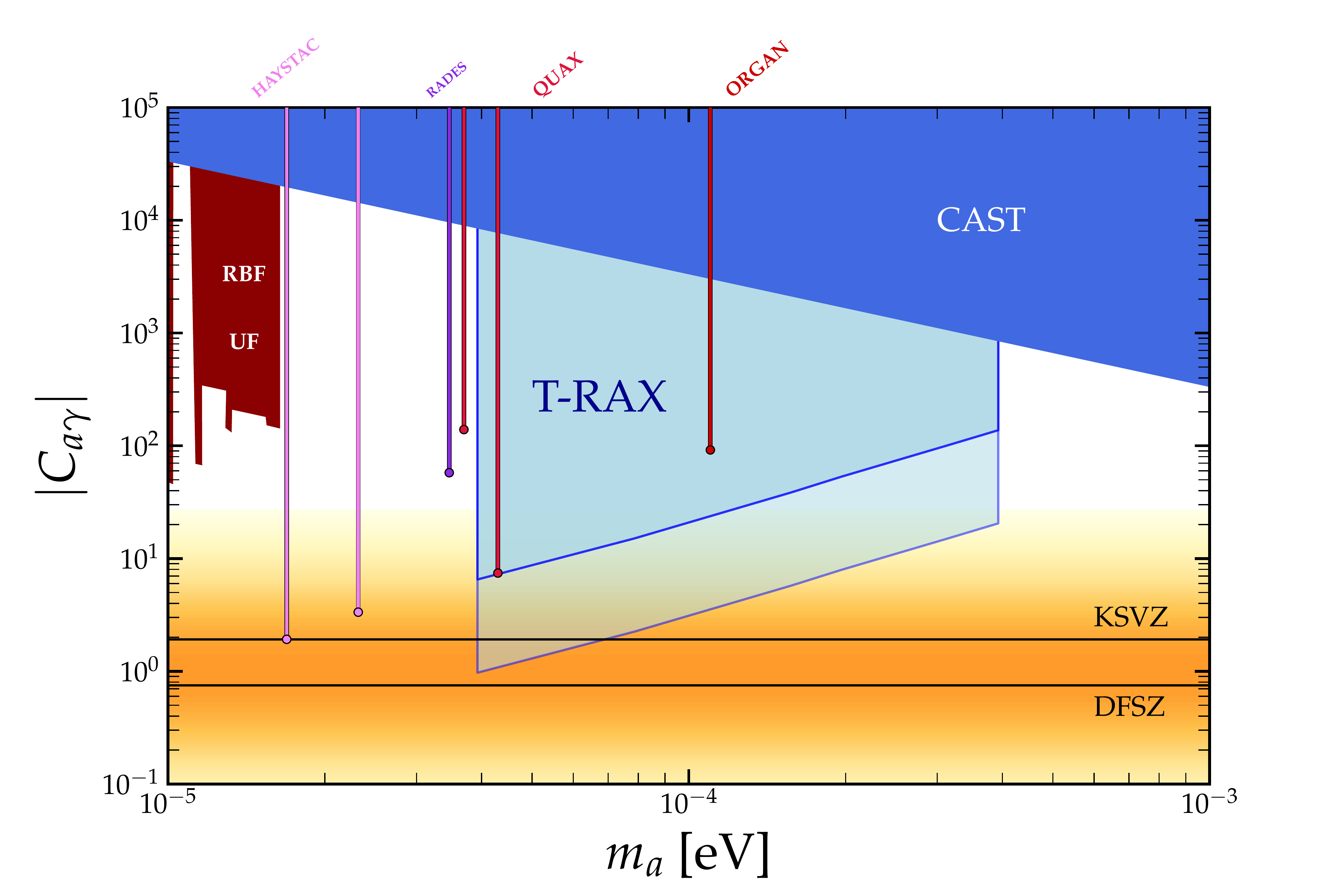}
\caption{\label{fig:sensitivity} Projected sensitivity of T-RAX for axion-photon coupling. The upper section corresponds to the simulated geometry with the peak power boost around 80,000. The lower shade corresponds to the case with 500\,mm height and a 30-T magnetic field. The figure also shows limits from CAST~\cite{CAST:2017uph}, RBF-UF~\cite{DePanfilis, Hagmann}, HAYSTACK~\cite{HAYSTAC:2018rwy,HAYSTAC:2020kwv}, RADES~\cite{CAST:2021add}, and ORGAN~\cite{MCALLISTER201767}. Image created using AxionLimits~\cite{ciaran_o_hare_2020_3932430}.}
\end{figure}

Fig.~\ref{fig:sensitivity} shows the possible search range with T-RAX. We scale Eq.~\ref{eq:sensitivity} in the 10--100\,GHz range. We assume that a heterodyne receiver at the single quantum limit dominates $T_\text{sys}$. The sensitivity decreases with higher frequency because the detector width ($\propto \nu^{-1}$) and the power boost ($\propto \nu^{-\frac{2}{3}}$) decrease while $T_\text{sys}$ ($\propto \nu$) increases at high mass. The default geometry can reach $|C_{a\gamma}| = 6.5$. The sensitivity can improve with a taller detector and either the application of a superconductor~\cite{ahn2020superconducting,9699394} or the proposed strong magnet. The lower line indicates the sensitivity of a 500\,mm-tall system inside a 30-T magnetic field. It has a minimum $|C_{a\gamma}| = 1.4$. 

The projected scan rate is about 1\,MHz per day, counting only the net measurement time. We need an array of T-RAX to cover the wide mass range. To cover the whole 10--100\,GHz range, we need about a hundred fixed-width T-RAX with a 2.5\,\% relative bandwidth. The net measurement time is five hundred days without considering integration, cool down, or warm-up times. To speed up the scan, we can operate an array of T-RAX simultaneously in a magnet, especially at the high frequency where the detector size is smaller. For comparison, 17 variable-width T-RAX (15\,\% relative bandwidth) can scan the same range, but their details depend more heavily on the available technology and implementation. 

Finally, the application of advanced detection methods is promising for T-RAX~\cite{axion_squeezing,8395076,instruments5030025,PhysRevLett.126.141302}. T-RAX's taper can focus large-signal power into a single port. Thermal noise is negligible in this frequency range at millikelvin temperature and only the detector's intrinsic noise matters. As a result, a large signal-to-noise ratio is achievable. The advanced detection methods will significantly improve the sensitivity and the scanning speed. The details of the application are currently under study.

\section{Summary}\label{sec:conclusion}
We introduce a novel high-mass axion detector design. The detector consists of an elongated rectangular waveguide operating near its cutoff frequency. Dielectric layers inside the waveguide further enhance the axion-induced electric field via longitudinal resonances. A tapered waveguide collects a traveling wave signal that passes the dielectric layers. The structure instruments a large area with few parasitic high-order modes. The detector has the potential to scan the QCD-axion in a wide mass range from $40-400\,\mu$eV. The system is small enough to fit inside a solenoid and is relatively simple and easy to build with existing technology.

\acknowledgments
The authors thank X. Li, A. E. Ivanov, and S. Heyminck for the helpful discussion. The authors also thank D. Strom and A. Caldwell for revising English.


\newpage

\appendix

\section{Loss calculation}
This section calculates the loss inside a simple, single-dielectric T-RAX in Sec.~\ref{sec:1D_DH}. The standing wave inside T-RAX is similar to a rectangular TE101 or circular TM010 mode, except that $k_x$ is slightly larger than the cutoff wavenumber. We borrow the textbook description~\cite{Pozar:882338} and list the relevant parameters here. We point out that critical coupling may not necessarily optimize the signal-to-noise ratio of T-RAX or a dielectric haloscope due to their complex internal structure. The work to optimize the signal-to-noise ratio is ongoing.


First, we calculate the conductor loss $P_c$. We assume that the field distribution is approximately TE101, where resistive energy dissipation on the waveguide walls dominates the loss. 
\begin{equation}
    P_c = \frac{R_s E_R^2 \lambda^2}{2 \eta^2} \Big(\frac{l^2 ab}{d^2} + \frac{bd}{a^2} + \frac{l^2a}{2d} + \frac{d}{2a} \Big)
\end{equation}
In our geometry, $a \ll d \ll b$, and the dominant loss occurs on the E-plane walls. This simplifies the expression.
\begin{equation}
    P_c \approx \frac{R_s E_R^2 \lambda^2}{2 \eta^2} \frac{b d}{a^2}
\end{equation}
Near the TE10 cutoff, $\frac{\lambda}{a} \to \frac{1}{2}$. $\eta$ is the intrinsic impedance of the filling material. For air, $\eta = Z_0 \approx 377\,\Omega$. $R_s$ is the frequency-dependent sheet resistance of the waveguide wall. For copper at room temperature, $R_s = 0.035\,\Omega$ at 19\,GHz. 
\begin{equation}
    P_c \approx \frac{R_s }{8 Z_0^2} E_R^2 bd
\end{equation}

Using $P_c$, we calculate the unloaded quality factor $Q_0$. The unloaded quality factor of a TE$_{10l}$ mode inside a rectangular cavity is
\begin{equation}
\begin{split}
    Q_0 &= \frac{(k a d)^3 b \eta}{2 \pi^2 R_s} \frac{1}{(2 l^2 a^3 b + 2 b d^3 + l^2 a^3  d + a d^3)} \\
    &\approx \frac{(k a d)^3 b \eta}{2 \pi^2 R_s} \frac{1}{2bd^3} = \frac{(k a )^3 \eta}{4 \pi^2 R_s}.
\end{split}
\end{equation}
Near the TE10 mode cutoff, $k_xa \to \pi$, and
\begin{equation}
   \lim_{k_xa \to \pi} Q_0 = \frac{\pi^3 Z_0}{4 \pi^2 R_s} = \frac{\pi}{4} \frac{Z_0}{R_s} \approx 8,000.
\end{equation}
Cryogenic temperature and superconductor can significantly reduce $R_s$ and increases $Q_0$. Increasing frequency $\nu$ decreases $Q_0$ as $\nu^{-2/3}$ due to anomalous skin effect.

\section{Dielectric position calculation}\label{ch:dispersion}
The dielectric haloscope's dispersion determines the resonance condition. Unfortunately, wave propagation calculation through a dispersive system quickly becomes complicated. 

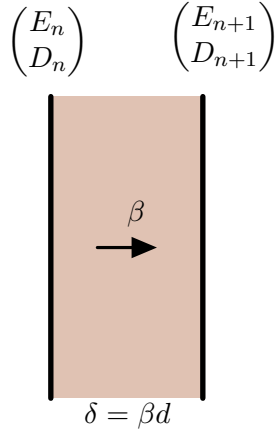
\begin{figure}
\begin{center}
\definecolor{zzttqq}{rgb}{0.6,0.2,0.}
\begin{tikzpicture}[line cap=round,line join=round,>=triangle 45,x=1.0cm,y=1.0cm]
\clip(-1.1104385594858852,-0.7642524989695316) rectangle (3.0505315623711993,5.426043215104716);
\fill[line width=0.pt,color=zzttqq,fill=zzttqq,fill opacity=0.30000001192092896] (0.,4.) -- (0.,0.) -- (2.,0.) -- (2.,4.) -- cycle;
\draw [line width=2.pt] (0.,0.)-- (0.,4.);
\draw [line width=2.pt] (2.,0.)-- (2.,4.);
\draw (0.30497340819500834,0.07135215664930619) node[anchor=north west] {$\delta = \beta d$};
\draw (-0.7182159660321437,5.34077743391912) node[anchor=north west] {$\begin{pmatrix} E_n \\ D_n  \end{pmatrix}$};
\draw (1.4134285636077564,5.408990058867597) node[anchor=north west] {$\begin{pmatrix} E_{n+1} \\ D_{n+1}     \end{pmatrix}$};
\draw [->,line width=1.2pt] (0.6119302204631532,1.9983588114437685) -- (1.396375407370636,1.9983588114437685);
\draw (0.8677275640199419,2.748697685877011) node[anchor=north west] {$\beta$};
\end{tikzpicture}
\caption{\label{fig:lekner_layer} Wave propagating through a transmission line}
\end{center}
\end{figure}

We use matrix formalism that relates the electric field and its derivative along the propagation direction~\cite{Lekner}. We prefer this formalism because all elements are real numbers in lossless systems, and its serial concatenation is a simple matrix multiplication of its constituent matrices. Fig.~\ref{fig:lekner_layer} shows a wave traveling through a transmission line section with a propagation constant $\beta$ and a physical thickness $d$. We use an electrical length $\delta = \beta d$ for simplicity. A matrix $L_n$ relates a traveling wave's electric field $E$ and its spatial derivative $D = \frac{d E}{d z}$ at left (subscript $n$) and right (subscript $n+1$) the transmission line section:
\begin{equation}
    \begin{pmatrix}
E_{n+1} \\
D_{n+1}
    \end{pmatrix} =
L_n
\begin{pmatrix}
E_n \\
D_n 
\end{pmatrix},
\end{equation}
where
\begin{equation}
    L_n = \begin{pmatrix} 
    \cos{\delta} & \frac{\sin{\delta}}{\beta} \\
    -\beta \sin{\delta} & \cos{\delta} 
\end{pmatrix}.\\
\end{equation}

We express one-dielectric T-RAX as 
\begin{equation}
\begin{split}
    \begin{pmatrix}
        E_o \\
        D_o 
    \end{pmatrix} &= L_d L_v \begin{pmatrix}
    0 \\
    1 
    \end{pmatrix}\\
    &= \begin{pmatrix} 
    \cos{\delta_d} & \frac{\sin{\delta_d}}{\beta_d} \\
    -\beta_d \sin{\delta_d} & \cos{\delta_d} 
\end{pmatrix} \begin{pmatrix} 
    \cos{\delta_v} & \frac{\sin{\delta_v}}{\beta_v} \\
    -\beta_v \sin{\delta_v} & \cos{\delta_v} 
\end{pmatrix} \begin{pmatrix}
    0 \\
    1 
    \end{pmatrix}\\
    &=\begin{pmatrix}
    \frac{\cos{\delta_v}\sin{\delta_d}}{\beta_d} + \frac{\cos{\delta_d}\sin{\delta_v}}{\beta_v} \\
    \cos{\delta_d}\cos{\delta_v} - \frac{\beta_d}{\beta_v}\sin{\delta_v}\sin{\delta_d} 
    \end{pmatrix}.\\
    \label{eq:l_matrix_t_rax_one}
\end{split}
\end{equation}
Subscripts $v$ and $d$ indicate vacuum and dielectric-filled waveguide sections. The last vector, $\begin{pmatrix}
    0 \\
    1 
    \end{pmatrix}$, is the boundary condition at the metallic mirror, \emph{i.e.} zero electric field and a unit electric field gradient. $\begin{pmatrix}
        E_o \\
        D_o 
    \end{pmatrix}$ is the traveling wave exiting the system at the last dielectric surface.

A similar equation below describes wave propagation inside a two-dielectric T-RAX. For simplicity, we use an identical electrical length $\delta_v$ among the mirror, the first dielectric, and the second dielectric. 
\begin{equation}
    \begin{pmatrix}
        E_o \\
        D_o 
    \end{pmatrix} = L_d L_v L_d L_v \begin{pmatrix}
    0 \\
    1 
    \end{pmatrix}
    \label{eq:l_matrix_t_rax}
\end{equation}

Resonances occur at $D_o = 0$, \emph{i.e.} when the one-way accrued phase is $\pm\frac{\pi}{2}$. The total phase accrued during a round trip and a reflection on the mirror is $2\pi$. Given fixed $\delta_d$, $\beta_d$, and $\beta_v$, we want to find $\delta_v$ near $\pi$ that makes $D_o = 0$. Below is an analytic solution for the single-dielectric case:
\begin{equation}
    \delta_v = \tan^{-1}{\Big(\frac{\beta_v}{\beta_d\tan{\delta_d}}\Big)}.
\end{equation}
For the two-dielectric case, we can find the solution numerically. A second-order Taylor expansion of $D_o$ with $\delta_v$ around $\pi$ is a good approximation.
\begin{equation}
\begin{split}
    D_o =& \cos{2\delta_d} - \frac{(3\beta_d^2 + \beta_v^2) \cos{\delta_d} \sin{\delta_d}}{\beta_v \beta_d}(\delta_v - \pi)\\ 
    &+ \Big( -\cos^2{\delta_d} -\cos{2\delta_d} + (\frac{\beta_d}{\beta_v}\sin{\delta_d})^2 \Big) (\delta_v - \pi)^2 + \mathcal{O}(\delta_v - \pi)^3 + ...
\end{split}
\end{equation}
Two $\delta_v$ satisfy $D_0=0$. The larger one, the symmetric resonance case, maximizes the power boost.

\bibliographystyle{JHEP}
\bibliography{trax}

\providecommand{\href}[2]{#2}\begingroup\raggedright\begin{thebibliography}{10}

\bibitem{Planck2018}
{Planck Collaboration}, {Aghanim, N.}, {Akrami, Y.}, {Ashdown, M.}, {Aumont,
  J.}, {Baccigalupi, C.} et~al., \emph{Planck 2018 results - vi. cosmological
  parameters}, \href{https://doi.org/10.1051/0004-6361/201833910}{\emph{A\&A}
  {\bfseries 641} (2020) A6}.

\bibitem{PhysRevD.16.1791}
R.D.~Peccei and H.R.~Quinn, \emph{Constraints imposed by $\mathrm{CP}$
  conservation in the presence of pseudoparticles},
  \href{https://doi.org/10.1103/PhysRevD.16.1791}{\emph{Phys. Rev. D}
  {\bfseries 16} (1977) 1791}.

\bibitem{PhysRevLett.38.1440}
R.D.~Peccei and H.R.~Quinn, \emph{$\mathrm{CP}$ conservation in the presence of
  pseudoparticles},
  \href{https://doi.org/10.1103/PhysRevLett.38.1440}{\emph{Phys. Rev. Lett.}
  {\bfseries 38} (1977) 1440}.

\bibitem{PhysRevLett.40.223}
S.~Weinberg, \emph{A new light boson?},
  \href{https://doi.org/10.1103/PhysRevLett.40.223}{\emph{Phys. Rev. Lett.}
  {\bfseries 40} (1978) 223}.

\bibitem{PhysRevLett.40.279}
F.~Wilczek, \emph{Problem of strong $p$ and $t$ invariance in the presence of
  instantons}, \href{https://doi.org/10.1103/PhysRevLett.40.279}{\emph{Phys.
  Rev. Lett.} {\bfseries 40} (1978) 279}.

\bibitem{PRESKILL1983127}
J.~Preskill, M.B.~Wise and F.~Wilczek, \emph{Cosmology of the invisible axion},
  \href{https://doi.org/https://doi.org/10.1016/0370-2693(83)90637-8}{\emph{Physics
  Letters B} {\bfseries 120} (1983) 127}.

\bibitem{ABBOTT1983133}
L.~Abbott and P.~Sikivie, \emph{A cosmological bound on the invisible axion},
  \href{https://doi.org/https://doi.org/10.1016/0370-2693(83)90638-X}{\emph{Physics
  Letters B} {\bfseries 120} (1983) 133}.

\bibitem{DINE1983137}
M.~Dine and W.~Fischler, \emph{The not-so-harmless axion},
  \href{https://doi.org/https://doi.org/10.1016/0370-2693(83)90639-1}{\emph{Physics
  Letters B} {\bfseries 120} (1983) 137}.

\bibitem{Borsayini}
S.~Borsanyi, Z.~Fodor, J.~Guenther, K.H.~Kampert, S.D.~Katz, T.~Kawanai et~al.,
  \emph{Calculation of the axion mass based on high-temperature lattice quantum
  chromodynamics}, \href{https://doi.org/10.1038/nature20115}{\emph{Nature}
  {\bfseries 539} (2016) 69}.

\bibitem{Dine:2017swf}
M.~Dine, P.~Draper, L.~Stephenson-Haskins and D.~Xu, \emph{{Axions, Instantons,
  and the Lattice}},
  \href{https://doi.org/10.1103/PhysRevD.96.095001}{\emph{Phys. Rev. D}
  {\bfseries 96} (2017) 095001}
  [\href{https://arxiv.org/abs/1705.00676}{{\ttfamily 1705.00676}}].

\bibitem{Klaer:2017ond}
V.B..~Klaer and G.D.~Moore, \emph{{The dark-matter axion mass}},
  \href{https://doi.org/10.1088/1475-7516/2017/11/049}{\emph{JCAP} {\bfseries
  11} (2017) 049} [\href{https://arxiv.org/abs/1708.07521}{{\ttfamily
  1708.07521}}].

\bibitem{Buschmann:2019icd}
M.~Buschmann, J.W.~Foster and B.R.~Safdi, \emph{{Early-Universe Simulations of
  the Cosmological Axion}},
  \href{https://doi.org/10.1103/PhysRevLett.124.161103}{\emph{Phys. Rev. Lett.}
  {\bfseries 124} (2020) 161103}
  [\href{https://arxiv.org/abs/1906.00967}{{\ttfamily 1906.00967}}].

\bibitem{Buschmann:2021sdq}
M.~Buschmann, J.W.~Foster, A.~Hook, A.~Peterson, D.E.~Willcox, W.~Zhang et~al.,
  \emph{Dark matter from axion strings with adaptive mesh refinement},
  \href{https://doi.org/10.1038/s41467-022-28669-y}{\emph{Nature
  Communications} {\bfseries 13} (2022) 1049}.

\bibitem{Ballesteros:2016euj}
G.~Ballesteros, J.~Redondo, A.~Ringwald and C.~Tamarit, \emph{{Unifying
  inflation with the axion, dark matter, baryogenesis and the seesaw
  mechanism}},
  \href{https://doi.org/10.1103/PhysRevLett.118.071802}{\emph{Phys. Rev. Lett.}
  {\bfseries 118} (2017) 071802}
  [\href{https://arxiv.org/abs/1608.05414}{{\ttfamily 1608.05414}}].

\bibitem{PhysRevD.32.2988}
P.~Sikivie, \emph{Detection rates for ``invisible''-axion searches},
  \href{https://doi.org/10.1103/PhysRevD.32.2988}{\emph{Phys. Rev. D}
  {\bfseries 32} (1985) 2988}.

\bibitem{PhysRevD.36.974}
P.~Sikivie, \emph{Erratum: Detection rates for "invisible"-axion searches},
  \href{https://doi.org/10.1103/PhysRevD.36.974}{\emph{Phys. Rev. D} {\bfseries
  36} (1987) 974}.

\bibitem{PhysRevLett.120.151301}
{\scshape ADMX Collaboration} collaboration, \emph{Search for invisible axion
  dark matter with the axion dark matter experiment},
  \href{https://doi.org/10.1103/PhysRevLett.120.151301}{\emph{Phys. Rev. Lett.}
  {\bfseries 120} (2018) 151301}.

\bibitem{PhysRevLett.118.061302}
B.M.~Brubaker et~al., \emph{First results from a microwave cavity axion search
  at $24\text{ }\text{ }\ensuremath{\mu}\mathrm{eV}$},
  \href{https://doi.org/10.1103/PhysRevLett.118.061302}{\emph{Phys. Rev. Lett.}
  {\bfseries 118} (2017) 061302}.

\bibitem{PhysRevLett.118.091801}
{\scshape MADMAX Working Group} collaboration, \emph{Dielectric haloscopes: A
  new way to detect axion dark matter},
  \href{https://doi.org/10.1103/PhysRevLett.118.091801}{\emph{Phys. Rev. Lett.}
  {\bfseries 118} (2017) 091801}.

\bibitem{orpheus}
R.~Cervantes, G.~Carosi, C.~Hanretty, S.~Kimes, B.H.~LaRoque, G.~Leum et~al.,
  \emph{Admx-orpheus first search for 70 $\mu$ev dark photon dark matter:
  Detailed design, operations, and analysis},  2022.
\newblock 10.48550/ARXIV.2204.09475.

\bibitem{Kuo_2020}
C.-L.~Kuo, \emph{Large-volume centimeter-wave cavities for axion searches},
  \href{https://doi.org/10.1088/1475-7516/2020/06/010}{\emph{Journal of
  Cosmology and Astroparticle Physics} {\bfseries 2020} (2020) 010}.

\bibitem{Kuo_2021}
C.-L.~Kuo, \emph{Symmetrically tuned large-volume conic shell-cavities for
  axion searches},
  \href{https://doi.org/10.1088/1475-7516/2021/02/018}{\emph{Journal of
  Cosmology and Astroparticle Physics} {\bfseries 2021} (2021) 018}.

\bibitem{Melc_n_2018}
A.{\'{A}}.~Melc{\'{o}}n, S.A.~Cuendis, C.~Cogollos, A.~D{\'{\i}}az-Morcillo,
  B.~Döbrich, J.D.~Gallego et~al., \emph{Axion searches with microwave
  filters: the {RADES} project},
  \href{https://doi.org/10.1088/1475-7516/2018/05/040}{\emph{Journal of
  Cosmology and Astroparticle Physics} {\bfseries 2018} (2018) 040}.

\bibitem{1129671}
L.~Lewin, \emph{The e-plane taper junction in rectangular waveguide},
  \href{https://doi.org/10.1109/TMTT.1979.1129671}{\emph{IEEE Transactions on
  Microwave Theory and Techniques} {\bfseries 27} (1979) 560}.

\bibitem{ParabolicTaper}
J.L.~Doane, \emph{Parabolic tapers for overmoded waveguides},
  \href{https://doi.org/10.1007/BF01009605}{\emph{International Journal of
  Infrared and Millimeter Waves} {\bfseries 5} (1984) 737}.

\bibitem{Warden2012CryogenicNF}
R.M.~Warden, \emph{Cryogenic nano-actuator for jwst},  in \emph{Proc. of the
  38th Aerospace Mechanisms Symposium}, 2006.

\bibitem{https://doi.org/10.48550/arxiv.2203.14923}
J.~Jaeckel, G.~Rybka and L.~Winslow, \emph{Axion dark matter},  2022.
\newblock 10.48550/ARXIV.2203.14923.

\bibitem{10.5555/27077}
N.~Marcuvitz, \emph{Waveguide Handbook}, Institution of Electrical Engineers,
  GBR (1986).

\bibitem{wg_bend}
``Waveguide bends \& twists.''
  \url{http://mdllab.com/wp-content/themes/divi-child/pdf/Waveguide-Bends-Twists/waveguidebends_twists.pdf}.

\bibitem{10.1007/978-3-030-43761-9_2}
M.D.~Bird, \emph{Ultra-high field solenoids and axion detection},  in
  \emph{Microwave Cavities and Detectors for Axion Research}, G.~Carosi and
  G.~Rybka, eds., (Cham), pp.~9--16, Springer International Publishing, 2020.

\bibitem{ahn2020superconducting}
D.~Ahn, O.~Kwon, W.~Chung, W.~Jang, D.~Lee, J.~Lee et~al.,
  \emph{Superconducting cavity in a high magnetic field},  2020.

\bibitem{9699394}
J.~Golm, S.~Arguedas~Cuendis, S.~Calatroni, B.~Dobrich, C.~Cogollos,
  J.D.~Gallego~Puyol et~al., \emph{Thin film (high temperature) superconducting
  radiofrequency cavities for the search of axion dark matter},
  \href{https://doi.org/10.1109/TASC.2022.3147741}{\emph{IEEE Transactions on
  Applied Superconductivity} (2022) 1}.

\bibitem{axion_squeezing}
K.M.~Backes, D.A.~Palken, S.A.~Kenany, B.M.~Brubaker, S.B.~Cahn, A.~Droster
  et~al., \emph{A quantum enhanced search for dark matter axions},
  \href{https://doi.org/10.1038/s41586-021-03226-7}{\emph{Nature} {\bfseries
  590} (2021) 238}.

\bibitem{8395076}
L.S.~Kuzmin, A.S.~Sobolev, C.~Gatti, D.~Di~Gioacchino, N.~Crescini, A.~Gordeeva
  et~al., \emph{Single photon counter based on a josephson junction at 14 ghz
  for searching galactic axions},
  \href{https://doi.org/10.1109/TASC.2018.2850019}{\emph{IEEE Transactions on
  Applied Superconductivity} {\bfseries 28} (2018) 1}.

\bibitem{instruments5030025}
A.~Rettaroli, D.~Alesini, D.~Babusci, C.~Barone, B.~Buonomo, M.M.~Beretta
  et~al., \emph{Josephson junctions as single microwave photon counters:
  Simulation and characterization},
  \href{https://doi.org/10.3390/instruments5030025}{\emph{Instruments}
  {\bfseries 5} (2021) }.

\bibitem{PhysRevLett.126.141302}
A.V.~Dixit, S.~Chakram, K.~He, A.~Agrawal, R.K.~Naik, D.I.~Schuster et~al.,
  \emph{Searching for dark matter with a superconducting qubit},
  \href{https://doi.org/10.1103/PhysRevLett.126.141302}{\emph{Phys. Rev. Lett.}
  {\bfseries 126} (2021) 141302}.

\bibitem{CAST:2017uph}
{\scshape CAST} collaboration, \emph{{New CAST Limit on the Axion-Photon
  Interaction}}, \href{https://doi.org/10.1038/nphys4109}{\emph{Nature Phys.}
  {\bfseries 13} (2017) 584}
  [\href{https://arxiv.org/abs/1705.02290}{{\ttfamily 1705.02290}}].

\bibitem{DePanfilis}
S.~DePanfilis, A.C.~Melissinos, B.E.~Moskowitz, J.T.~Rogers, Y.K.~Semertzidis,
  W.U.~Wuensch et~al., \emph{Limits on the abundance and coupling of cosmic
  axions at 4.5$<{m}_{a}<$5.0 \ensuremath{\mu}ev},
  \href{https://doi.org/10.1103/PhysRevLett.59.839}{\emph{Phys. Rev. Lett.}
  {\bfseries 59} (1987) 839}.

\bibitem{Hagmann}
C.~Hagmann, P.~Sikivie, N.S.~Sullivan and D.B.~Tanner, \emph{Results from a
  search for cosmic axions},
  \href{https://doi.org/10.1103/PhysRevD.42.1297}{\emph{Phys. Rev. D}
  {\bfseries 42} (1990) 1297}.

\bibitem{HAYSTAC:2018rwy}
{\scshape HAYSTAC} collaboration, \emph{{Results from phase 1 of the HAYSTAC
  microwave cavity axion experiment}},
  \href{https://doi.org/10.1103/PhysRevD.97.092001}{\emph{Phys. Rev. D}
  {\bfseries 97} (2018) 092001}
  [\href{https://arxiv.org/abs/1803.03690}{{\ttfamily 1803.03690}}].

\bibitem{HAYSTAC:2020kwv}
{\scshape HAYSTAC} collaboration, \emph{{A quantum-enhanced search for dark
  matter axions}},
  \href{https://doi.org/10.1038/s41586-021-03226-7}{\emph{Nature} {\bfseries
  590} (2021) 238} [\href{https://arxiv.org/abs/2008.01853}{{\ttfamily
  2008.01853}}].

\bibitem{CAST:2021add}
{\scshape CAST} collaboration, \emph{{First results of the CAST-RADES haloscope
  search for axions at 34.67 $\mu$eV}},
  \href{https://arxiv.org/abs/2104.13798}{{\ttfamily 2104.13798}}.

\bibitem{MCALLISTER201767}
B.T.~McAllister, G.~Flower, E.N.~Ivanov, M.~Goryachev, J.~Bourhill and
  M.E.~Tobar, \emph{The organ experiment: An axion haloscope above 15 ghz},
  \href{https://doi.org/https://doi.org/10.1016/j.dark.2017.09.010}{\emph{Physics
  of the Dark Universe} {\bfseries 18} (2017) 67}.

\bibitem{ciaran_o_hare_2020_3932430}
C.~O'Hare, \emph{cajohare/axionlimits: Axionlimits},  July, 2020.
\newblock 10.5281/zenodo.3932430.

\bibitem{Pozar:882338}
D.M.~Pozar, \emph{{Microwave engineering; 3rd ed.}}, Wiley, Hoboken, NJ (2005).

\bibitem{Lekner}
J.~Lekner, \emph{Matrix and numerical methods},  in \emph{{Theory of Reflection
  of Electromagnetic and Particle Waves}}, pp.~224--228, Springer (1987),
  \href{https://link.springer.com/book/10.1007/978-3-319-23627-8}{https://link.springer.com/book/10.1007/978-3-319-23627-8}.

\end{thebibliography}\endgroup
\end{document}